\documentclass[12pt]{article}
\usepackage{epsfig}
\usepackage{amssymb,amsmath}
\usepackage{setspace}
\pdfoutput=1
\usepackage{subfigure}
\usepackage{amssymb,amsmath}
\usepackage{graphicx}
\usepackage{color}
\usepackage{cancel}
\usepackage[colorlinks=true
,urlcolor=blue
,citecolor=blue
,linkcolor=blue
,pagecolor=blue
,linktocpage=true
,pdfproducer=medialab
]{hyperref}
\def\bi{\begin{itemize}}
\def\ei{\end{itemize}}

\def\ttau{\tilde \tau}

\def\alt{\lesssim}
\def\agt{\gtrsim}

\newcommand\jhep[3]{{\it J. High Energy Phys.\ }{\bf #1} (#2) #3}

\def\ttau{\tilde \tau}

\usepackage{amsmath}

\newcommand{\bea}{\begin{eqnarray}}
\newcommand{\eea}{\end{eqnarray}}

\newcommand{\beq}{\begin{equation}}
\newcommand{\eeq}{\end{equation}}

 \makeatletter
\def\alt{\mathrel{\mathpalette\gl@align<}}
\def\agt{\mathrel{\mathpalette\gl@align>}}
\def\gl@align#1#2{\lower.6ex\vbox{\baselineskip\z@skip\lineskip\z@
\ialign{$\m@th#1\hfil##\hfil$\crcr#2\crcr\sim\crcr}}} \makeatother

\usepackage{longtable}

\begin{document}
\begin{flushright}
ACT-02-15\\
\end{flushright}

\vspace*{1.0cm}

\begin{center}
\baselineskip 20pt {\Large\bf
Super-Natural MSSM
}
\vspace{1cm}

{\large
Guangle Du$^{a,}$\footnote{E-mail:gldu@itp.ac.cn},
Tianjun Li$^{a,b,}$\footnote{E-mail:tli@itp.ac.cn},
D.V. Nanopoulos$^{c,d,}$\footnote{E-mail:dimitri@physics.tamu.edu},
Shabbar Raza$^{a,}$\footnote{E-mail:shabbar@itp.ac.cn}, 
} \vspace{.5cm}

{\baselineskip 20pt \it $^a$
State Key Laboratory of Theoretical Physics and Kavli Institute for Theoretical Physics China (KITPC),
Institute of Theoretical Physics, Chinese Academy of Sciences, Beijing 100190, P. R. China \\
}
{\it $^b$
School of Physical Electronics, University of Electronic Science and Technology of China,\\
Chengdu 610054, P. R. China \\
}

{\it $^c$
George P. and Cynthia W. Mitchell Institute for Fundamental Physics, Texas A$\&$M University, College Station, Texas 77843, USA\\
}
{\it $^d$
Astroparticle Physics Group, Houston Advanced Research Center (HARC), Mitchell Campus, Woodlands, Texas 77381, USA
and Academy of Athens, Division of Natural Sciences, 28 Panepistimiou Avenue, Athens 10679, Greece
}



\vspace{.5cm}

\vspace{1.5cm} {\bf Abstract}
\end{center}

We point out that the electroweak fine-tuning problem in the supersymmetric Standard Models (SSMs) is mainly
due to the high energy definition of the fine-tuning measure. We propose super-natural supersymmetry 
which has an order one high energy fine-tuning measure automatically. The key point is that
 all the mass parameters in the SSMs arise from a single supersymmetry breaking parameter. In this paper,
we show that there is no supersymmetry electroweak fine-tuning problem explicitly
in the Minimal SSM (MSSM) with no-scale supergravity and Giudice-Masiero (GM) mechanism.
We demonstrate that the $Z$-boson mass, the supersymmteric Higgs mixing parameter $\mu$ at the unification scale,
and the sparticle spectrum can be given as functions of the universal gaugino mass $M_{1/2}$. 
Because the light stau is the lightest supersymmetric particle (LSP) in the no-scale MSSM, 
to preserve $R$ parity, we introduce a non-thermally generated axino as the LSP dark matter candidate.
We estimate the lifetime of the light stau by calculating its 2-body and 3-body decays to 
the LSP axino for several values of axion decay constant $f_a$, and find that the light stau has
a lifetime $\tau_{\tilde \tau_1}$ in $[10^{-4},100]$ s for an $f_a$ range $[10^{9},10^{12}]$ GeV. We show
that our next to the LSP stau solutions are consistent with all the current experimental constraints,
including the sparticle mass bounds, B-physics bounds, Higgs mass, cosmological bounds,
 and the bounds on long-lived charge particles at the LHC.

\thispagestyle{empty}

\newpage

\addtocounter{page}{-1}

\baselineskip 18pt

\section{Introduction}

It is well-known that supersymmetry (SUSY) solves the gauge hierarchy problem in the Standard Model (SM) naturally.
From the low energy physics point of view, in the supersymmetric SMs (SSMs) with $R$-parity conservation, 
gauge coupling unification can be achieved, the Lightest Supersymmetric Particle (LSP) such as neutralino 
can be a dark matter (DM) candidate, and the electroweak (EW) gauge symmetry can be broken radiatively 
due to the large top quark Yukawa coupling, etc. While from the high energy physics point of view,
gauge coupling unification strongly suggests Grand Unified Theories (GUTs), 
and only the superstring theory may describe 
the real world. Therefore, supersymmetry is a bridge between the low energy phenomenology and high-energy
fundamental physics, and then is the most promising new physics beyond the SM.

A SM-like Higgs boson with mass $m_h$ around 125 GeV was discovered in July 2012 from 
the first run of the LHC~\cite{ATLAS, CMS}. This is indeed a little bit heavy in
the Minimal SSM (MSSM) since it requires the multi-TeV top squarks with small mixing or TeV-scale 
top squarks with large mixing~\cite{Carena:2011aa}. But it is fine in the next to MSSM (NMSSM) since
 the SM-like Higgs boson mass can be lifted further via
the extra tree-level contribution and pushing up effect from the Higgs mass matrix
diagonalization~\cite{Ellwanger:2011aa, Kang:2012sy, Cao:2012fz}. 
In addition, we have strong constraints on the parameter space in the SSMs from the LHC SUSY searches.
For instance, the gluino mass $m_{\tilde g}$ and first two-generation squark mass $m_{\tilde q}$  
should be heavier than about 1.7 TeV if they are roughly degenerate $m_{\tilde q} \sim m_{\tilde g}$, 
and the gluino mass is heavier than about 1.3 TeV for $m_{\tilde q} \gg m_{\tilde g}$~\cite{Aad:2014wea, Chatrchyan:2013wxa}.
Therefore, the naturalness in the SSMs is challenged from both the Higgs boson mass and the LHC supersymmetry searches.
As we know, there are two kinds of definitions for fine-tuning measures: 
the low energy definition~\cite{Kitano:2005wc, Papucci:2011wy, Baer:2012mv} and
high energy definition~\cite{Ellis:1986yg,Barbieri:1987fn}, which will be explained in details later. 
We shall point out that the naturalness conditions
from the low energy definition can still be satisfied in principle, but the naturalness condition
from the high energy definition is indeed a big challenge. However, supersymmetry is the connection
between the low and the high energy physics, so we should consider seriously the fine-tuning problem from the high
energy definition. Recently, it was found~\cite{Leggett:2014mza,Leggett:2014hha} that
the high energy fine-tuning measure~\cite{Ellis:1986yg,Barbieri:1987fn} 
will automatically be at the order one ${\cal O} (1)$ in the 
${\cal F}$-$SU(5)$ models~\cite{F-SU5, Jiang:2006hf, Jiang:2009zza, Li:2010ws} with 
the No-Scale supergravity (SUGRA) boundary conditions~\cite{Cremmer:1983bf}
and the Giudice-Masiero (GM) mechanism~\cite{Giudice:1988yz}.
Based on the above results, we propose the super-natural supersymmetry 
whose high energy fine-tuning measure is at the order one ${\cal O} (1)$ naturally.
The essential point for super-natural supersymmetry is that
all the mass parameters in the SSMs arise from a single supersymmetry breaking parameter.

In this paper, we show that there is no residual electroweak fine-tuning (EWFT) left in the MSSM 
if we employ the No-Scale supergravity boundary conditions~\cite{Cremmer:1983bf}
 and GM mechanism~\cite{Giudice:1988yz}. From no-scale supergravity, we have the 
supersymmetry breaking (SSB) soft terms: $M_{1/2}\not=0$ and $m_{0}=A_{0}=B_{0} = 0$,  where 
$M_{1/2}$,  $m_{0}$, $A_{0}$ and $B_{0}$ are respectively the universal 
gaugino mass, scalar mass, trilinear $A$ term, and Higgs mass mixing parameter at the GUT scale. 
We study various aspects of the No-Scale MSSM. We show how the parameter $c$, which is the ratio 
between the Higgs bilinear mass parameter $\mu(GUT)$ at the GUT scale and gaugino mass $M_{1/2}$, 
changes as a function of the $Z$-boson mass $M_Z$. In the no-scale MSSM, the 
light stau ($\tilde \tau_1$) turns out to be the LSP, which has a dominant $\tilde \tau_R$ component. 
To solve the cosmological dark matter problem and keep the $R$ parity intact,
we introduce a non-thermally generated axino ($\tilde a$) as the LSP dark matter candidate.
In other words, the light stau is the next to LSP (NLSP), and can decay to 
the LSP axino. The extremely weak interactions between axino and the lighter stau can cause the long-lived staus. 
The long-lived charged massive particles (sometimes also called as CHAMPs) like the stau 
in our case can disturb the successful predictions of Big Bang Nucleosynthesis (BBN), and may
 violate the other cosmological constraints such as 
catalyst BBN (CBBN) and structure formation bounds. Thus, we will estimate 
the lifetime of the NLSP stau by calculating 2-body ($\tilde \tau_{R}\rightarrow \tau \tilde a$) 
and 3-body ($\tilde \tau_{R}\rightarrow \tau \gamma \tilde a$) decays. In addition,  
in the viable parameter spaces which satisfy the sparticle mass bounds, B-physics bounds, 
and $m_{h}=125 \pm 2$ GeV, we obtain  $M_{1/2}\gtrsim$ 1800 GeV and $\tan\beta\gtrsim$ 27.
We find that the NLSP stau can have lifetime $\tau_{\tilde \tau_1}$ from $10^{-4}$~s to $100$~s
for the axion decay constant $f_a$ range $[10^{9},10^{12}]$ GeV, which is
 consistent with all the above mentioned cosmological bounds. If we consider $f_a=7\times 10^{8}$ ~\footnote{Various astrophysical and cosmological considerations and axion searches suggest the lower limit of axion decay
constant as $f_{a} \gtrsim 6\times 10^{8}$ GeV~\cite{Raffelt:2006cw}.}, 
we estimate lifetime $\tau_{\tilde \tau_1}$ from $10^{-7}$~s to $10^{-4}$~s depending upon $|e_{Q}|$ (will be defined latter in the text) values,
corresponding to the decay lengths of around a hundred meter to around 30 kilometers.
Moreover, our NLSP
stau satisfies the recently reported ATLAS Collaboration bounds on the long-lived staus as well. 
We also comment on the possible way of detection of staus in such a scenario. 
Furthermore, we present three benchmark points with gaugino masses about 1866~GeV,
2725~GeV, and 4589~GeV as well as Higgs boson masses about 123~GeV, 125~GeV,
and 127~GeV, respectively. The particle spectra are generically heavy due to large gaugino masses.

This paper is organized as follows. In Section~2, we briefly review the No-Scale supergravity. 
In Section~3, we discuss the supersymmetry electroweak fine-tuning problems and propose 
the super-natural supersymmetry. In Section~4, we present the numerical calculations of EWFT
and display various aspect of our study by presenting several graphs.
In Section~\ref{sec:dm}, we study dark matter in the No-Scale MSSM. We outline the detailed scanning procedure,
and the relevant experimental constraints that we have considered. Results of our scans are given as well.
A summary and conclusion are given in Section~\ref{summary}.

\section{Brief Review of No-Scale Supergravity}

Let us first briefly review the basic idea of No-Scale Supergravity, which 
was proposed to address the cosmological flatness problem~\cite{Cremmer:1983bf}. 
It fulfill three constraints: i) the vacuum energy vanishes automatically 
due to the appropriate K\"ahler potential; ii) There exist flat directions 
that leave the gravitino mass $M_{3/2}$ undetermined at the minimum of 
the scalar potential (this is why it is called as No-Scale model); 
iii) The quantity ${\rm Str} {\cal M}^2$ is zero at the minimum. 
The large one-loop corrections would force $M_{3/2}$ to be either identically 
zero or of the Planck scale in case of violation of 
the third condition. A minimal K\"ahler potential that satisfies the first 
two conditions is~\cite{Cremmer:1983bf}
\begin{eqnarray} 
K &=& -3 {\rm ln}( T+\overline{T}-\sum_i \overline{\Phi}_i
\Phi_i)~,~
\label{NS-Kahler}
\end{eqnarray}
where $T$ is a modulus field and $\Phi_i$ are matter fields. They parametrize the non-compact $SU(N,1)/SU(N) \times U(1)$ coset space. The third condition can always be satisfied in principle and is model dependent~\cite{Ferrara:1994kg}. From the K\"ahler potential in Eq.~(\ref{NS-Kahler}) one automatically obtains the No-Scale boundary conditions $m_0 = A_{0} = B_{0} = 0$, while $M_{1/2}$ can be non-zero and evolve naturally, as is in fact required for supersymmetry breaking. The high-energy boundary condition $B_{0} = 0$ effectively fixes $\tan\beta$ at low energy. 
 This means the entire supersymmetric particle (sparticle) spectrum is determined by $M_{1/2}$ and 
in a very good approximation the whole sparticle spectra are linearly rescaled in terms of $M_{1/2}$. 
The result is a natural $one$-$parameter$ model, with $M_{1/2}$ the single degree of freedom for mass parameter.


\section{The Supersymmetry Electroweak Fine-Tuning Problem and Super-Natural Supersymmetry}

To consider the fine-tuning issue in the supersymmetric SMs, we need to define the
fine-tuning measures first. There are two kinds of definitions: the low energy definition and high energy definition.

For low energy definition, there are two similar definitions. Let us review them one by one.
The first definition considers the Higgs boson mass~\cite{Kitano:2005wc, Papucci:2011wy}.
The SM-like Higgs particle $h$ in the MSSM is a linear combination
of $H_u^0$ and $H_d^0$.  Its potential can simply be reduced to
\begin{eqnarray}
V&=& \overline{m}^2_h |h|^2 + {{\lambda_h}\over 4} |h|^4~,~
\end{eqnarray}
where $\overline{m}^2_h$ is negative. Minimizing the Higgs potential,
we obtain the physical SM-like Higgs boson mass $m_h$
\begin{eqnarray}
m_h^2 = - 2 \overline{m}^2_h ~.~
\end{eqnarray}
The fine-tuning measure can be defined as~\cite{Kitano:2005wc}
\begin{eqnarray}
\Delta_{\rm FT} \equiv {{2 \delta \overline{m}^2_h}\over {m_h^2}}~.~
\end{eqnarray}

For a moderately large $\tan\beta\equiv \langle H_u^0 \rangle/\langle H_d^0 \rangle$,
for example, $\tan\beta \ge 2$, we have
\begin{eqnarray}
\overline{m}^2_h & \simeq & |\mu|^2 + m^2_{H_u}|_{\rm tree}+m^2_{H_u}|_{\rm rad} ~,~\,
\end{eqnarray}
where $\mu$ is the supersymmetric bilinear mass between $H_u$ and $H_d$, and
$m^2_{H_u}|_{\rm tree}$ and $m^2_{H_u}|_{\rm rad}$ are
the tree-level and radiative contributions to the soft supersymmetry-breaking
 mass squared for $H_u$.
Therefore, we obtain the following naturalness bounds for $5\%$ fine-tuning~\cite{Papucci:2011wy}
\begin{itemize}

\item The $\mu$ term or effective $\mu$ term is smaller than 400 GeV.

\item The square root  $M_{\tilde t} \equiv \sqrt{m_{{\tilde{t}_1}}^2+m_{\tilde{t}_2}^{2}}$
of the sum of the two stop mass squares
 is smaller than
1.2~TeV. Consequently, we can show that the light sbottom mass is lighter than
$m_{\tilde{t}_2}$.

\item The gluino mass is lighter than 1.8 TeV.

\end{itemize}

The second low energy definition, which is similar, considers the $Z$ boson mass. 
With the one-loop effective potential contributions to the tree-level MSSM 
Higgs potential, we get the $Z$-bosom mass $M_Z$
\begin{equation}
\frac{M_Z^2}{2} =
\frac{(m_{H_d}^2+\Sigma_d^d)-(m_{H_u}^2+\Sigma_u^u)\tan^2\beta}{\tan^2\beta
-1} -\mu^2 \; ,
\label{eq:mssmmu}
\end{equation}
where $\Sigma_u^u$ and  $\Sigma_d^d$ are the contributions arising from the one-loop effective potential 
defined in Ref.~\cite{Baer:2012mv} and $\tan\beta \equiv \frac{v_u}{v_d}$ is the ratio
of the two Higgs Vacuum Expectation values (VEVs). All parameters  
in Eq. (\ref{eq:mssmmu}) are defined at the electroweak scale $M_{EW}$.
To measure the EWFT, we define~\cite{Baer:2012mv}
\begin{eqnarray}
&& C_{H_d}\equiv |m_{H_d}^2/(\tan^2\beta -1)|,\,\, C_{H_u}\equiv
|-m_{H_u}^2\tan^2\beta /(\tan^2\beta -1)|, \, \, C_\mu\equiv |-\mu^2 |, \nonumber \\
&& C_{\Sigma_{d}^{d}} \equiv |\Sigma_d^d/(\tan^2\beta -1)|,\,\,
C_{\Sigma_{u}^{u}} \equiv |-\Sigma_u^u \tan^2\beta/(\tan^2\beta -1)|~.~
\label{cc1}
\end{eqnarray}
The fine-tuning measure is defined by~\cite{Baer:2012mv}
\begin{equation}
 \Delta_{\rm EW}\equiv {\rm max}(2C_k )/(M_Z^2)~.
\label{eq:ewft}
\end{equation}
Note that $\Delta_{EW}$ only depends on the weak-scale parameters of the SSMs, and then is fixed
by the particle spectra. Hence, it is independent of how the SUSY particle masses arise.

For high energy definition in the GUTs with gravity mediated supersymmetry breaking,
the typical quantitative measure $\Delta_{\rm EENZ}$ for fine-tuning is defined by
the maximum of the logarithmic derivative of $M_Z$ with respect to all fundamental parameters $a_i$ 
at the GUT scale~\cite{Ellis:1986yg,Barbieri:1987fn}
\begin{eqnarray}
\Delta_{\rm EENZ} ~=~ {\rm Max}\{\Delta_i^{\rm GUT}\}~,~~~
\Delta_i^{\rm GUT}~=~\left|
\frac{\partial{\rm ln}(M_Z)}{\partial {\rm ln}(a_i^{\rm GUT})}
\right|~.~\,
\label{eq:BG-EENZ}
\end{eqnarray}

Because the SM-like Higgs boson mass in the MSSM is smaller than the $Z$ boson mass at tree level 
and can be lifted by the top squarks radiatively, the discovery of Higgs boson with mass 
around 125 GeV~\cite{ATLAS,CMS} indeed constrains the viable parameter spaces in the SSMs. 
In the MSSM, we might need the large trilinear soft term $A_t$ or say large stop mixing if
we want stop masses around 1 TeV~\cite{Carena:2011aa}. In the NMSSM, the SM-like Higgs mass can also be lifted via
the additional tree-level contribution and pushing up effect from the Higgs mass matrix
diagonalization~\cite{Ellwanger:2011aa, Kang:2012sy, Cao:2012fz}. 
So the NMSSM looks more natural. In short, the 125 GeV Higgs boson mass
will not induce big fine-tuning in the SSMs from the low energy definitions of fine-tuning
measures. On the other hand, although the null results of the 
LHC Run1 have raised the lower bounds on the masses of gluino, first/second generation squarks,
and sleptons~\cite{Aad:2014wea,Chatrchyan:2013wxa,Aad:2014nua,Aad:2014vma,CMS-PAS-SUS-13-006},
they are still within the upper bounds from the $5\%$ fine-tuning requirements via
the low energy definitions given above. Therefore, the SSMs are still fine if we allow
a few percent fine-tuning from the low energy definitions. The key problem is the
high energy definition of fine-tuning measure. For example, we can have the benchmark points which have
the low energy fine-tuning measure $\Delta_{\rm EW}$ around 20 while the high energy fine-tuning
measure $\Delta_{\rm EENZ}$ around 1,500. For example, see the benchmark points 1 and 2
in Table 1 of Ref.~\cite{Li:2014dna}.

Because the fine-tuning measures for high energy definition in the viable SSMs are 
very large at the order of $10^3$ (${\cal O} (10^3)$), we would like to explore the supersymmetry breaking
scenario whose fine-tuning measure for high energy definition is automatically at
the order one (${\cal O} (1)$). In other word, the fine-tuning measure in Eq.~(\ref{eq:BG-EENZ})
is exactly one in the dream case. Interestingly, the solution 
with $\Delta_{\rm EENZ}=1$ is indeed simple. If there is one and only one mass parameter
$M_*$ in the SSMs, $M_{Z}$ is a trivial function of $M_*$, and we have the following approximate scale relation
\begin{eqnarray}
M_Z^n ~=~ f_n \left( c_i \right) ~M_*^n~,~\,
\label{eq:fn}
\end{eqnarray}
where $f_n$ is a dimensionless parameter, and $c_i$ denote the dimensionless coupling parameters,
such as gauge and Yukawa couplings, as well as the ratio between $\mu$ and $M_{1/2}$ in the MSSM, etc.

For the nearly constant $f_n$ of Eq. (\ref{eq:fn}), we have
\begin{eqnarray}
\frac{\partial M_Z^n}{\partial M_*^n} ~\simeq~ f_n~,~\, 
\label{eq:fnd}
\end{eqnarray}
 and therefore we obtain
\begin{eqnarray}
\frac{\partial{\rm ln}(M_Z^n)}{\partial {\rm ln}(M_{*}^n)}
~\simeq~  \frac{M_{*}^n}{M_Z^n} \frac{\partial M_Z^n}{\partial M_{*}^n}
~\simeq~  \frac{M_{*}^n}{M_Z^n} \frac{\delta M_Z^n}{\delta M_{*}^n} 
~\simeq~ \frac{1}{f_n} f_n~.~\
\label{fnpar}
\end{eqnarray}
Consequently, the fine-tuning measure is an order one constant
\begin{eqnarray}
\left|
\frac{\partial{\rm ln}(M_Z^n)}{\partial {\rm ln}(M_{*}^n)}
\right| \simeq {\cal O}(1)~.~\,
\label{eq:orderone}
\end{eqnarray}
 Therefore, there is no electroweak fine-tuning problem in such kind of SSMs. 
In the no-scale ${\cal F}$-$SU(5)$ model where the $\mu$ problem is solved via 
the GM mechanism~\cite{Giudice:1988yz}, we obtain  
$\mu \propto M_{1/2} \propto M_{3/2}$, suggesting mutual proportionality, 
where the gravitino mass can be situated around 30 TeV to elude the gravitino problem.
Crucially, finding $\mu \simeq M_{1/2}$ approximately rescales the sparticle spectra 
per variation only in $M_{1/2}$. Taking $M_*\equiv M_{1/2}$, it has been shown 
that there is indeed no residual EWFT problem~\cite{Leggett:2014mza,Leggett:2014hha}.


Based on the above discussions, we propose the super-natural supersymmetry with
 $\Delta_{\rm EENZ} \simeq 1$. The necessary conditions for the super-natural supersymmetry are

\begin{itemize}

\item {The K\"ahler potential and superpotential can be calculated in principle or at least inspired 
from a fundamental theory such as string theory with suitable compactifications. In other words, one cannot
add arbitrary high-dimensional terms in the K\"ahler potential and superpotential.  }

\item {There is one and only one chiral superfield or modulus which breaks
supersymmetry. And 
all the supersymmetry breaking soft terms are obtained from the above K\"ahler potential and superpotential.}

\item {All the other mass parameters, if there exist like the $\mu$ term in the MSSM,
 must arise from supersymmetry breaking.}

\end{itemize} 
Therefore, all the supersymmetry breaking soft terms and mass parameters in the SSMs 
are linearly proportional to the gravitino mass. 

In this paper we will show that there is no residual EWFT problem 
in the MSSM with the No-Scale SUGRA boundary conditions and GM mechanism.
 In order to achieve our goal, our
general strategy is as follows. With the gauge coupling unification, we shall
determine the GUT scale $M_{GUT}$. Then we obtain the ratio $\mu/M_{1/2}$,
top quark Yukawa coupling, and all remaining input parameters at $M_{GUT}$.
As a result, $M_{Z}$ is a trivial function of $M_{1/2}$.
Therefore, from the above discussions, there is no electroweak fine-tuning problem 
in such No-Scale MSSM. 
We shall confirm this in the following via numerical calculations of $\Delta_{\rm EENZ}$ for $n = 1$ 
in Eq.~(\ref{eq:orderone}) and point out that it is the 
same for the traditional choice of $n=2$ or any other positive integer $n$
since all the mass parameters are linearly proportional to 
$M_{1/2}$ in the No-Scale MSSM with the GM mechanism.


\section{The Numerical Calculations of $\Delta_{EENZ}$ in the No-Scale MSSM}
We use publicly available code SuSpect2.43~\cite{Djouadi:2002ze} for our calculations. We employ 
the no-scale SUGRA boundary conditions ($m_{0}=A_{0}=0$) and generate points randomly for the following parameter space 
\begin{align}
300\,\rm{GeV} \leq &   M_{1/2}   \leq 2000 \, \rm{GeV} ~,~\nonumber \\ 
1\leq & \tan\beta  \leq 60~.~
\label{input_param_range}
\end{align}

\begin{figure}[htp!]
\centering
\includegraphics[totalheight=6.0cm,width=8.0cm]{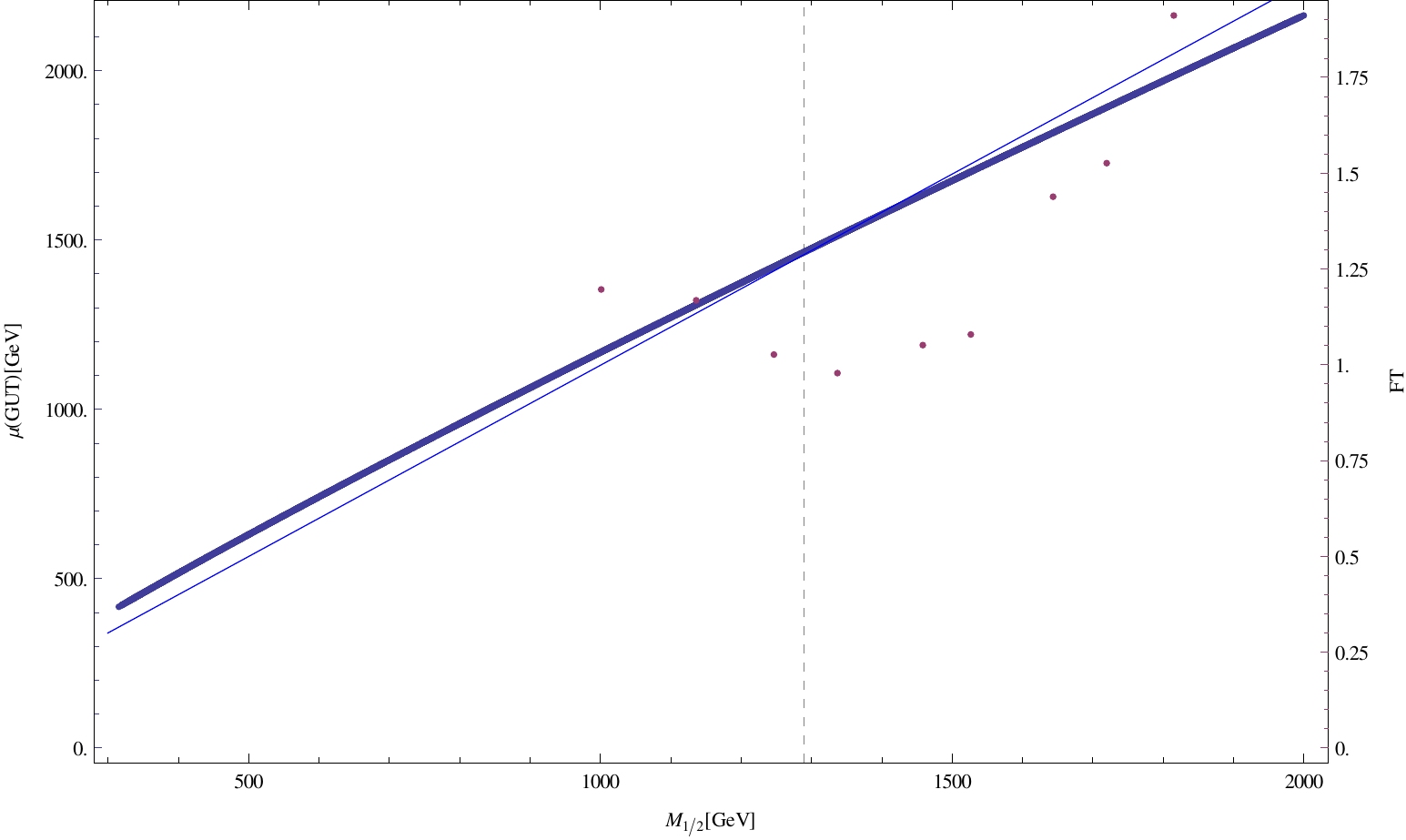}
\caption{
 $\mu(GUT)$ (left-vertical axis) and EWFT measure $\Delta_{EENZ}$ (right-vertical axis) versus $M_{1/2}$.
The thick blue line represents data generated, and the thin blue line is the line fitted to the data. 
Maroon points depict the values of $\Delta_{EENZ}$. 
}
\label{ft}
\end{figure}

We use ${\rm sign}(\mu)>0$, $m_t=173.3$ GeV~\cite{ATLAS:2014wva}  and $m_b=4.25$ GeV. We implement $B_{0}$ = 0 requirement numerically with a width $|B_{0}|\lesssim$ 1 GeV that
is comparable to the electroweak radiative corrections. In Fig.~\ref{ft}, 
 we plot  $\mu(GUT)$ and fine-tuning measure respectively on the left and the right vertical axes 
while $M_{1/2}$ is plotted on the horizontal axis. Let us discuss our calculations and Fig.~\ref{ft} in more details. 
We divide our calculations into two parts.\\
\\
$\bullet$ {\bf Part 1}: We vary $M_{1/2}$ and $\tan\beta$ in the interval given above Eq.~(\ref{input_param_range})
 with no-scale SUGRA boundary conditions and demand $|B_{0}|\lesssim$ 1 GeV. 
The thick blue line in Fig.~\ref{ft} is consisted of points we generated. Here we see that $\mu(GUT)\propto M_{1/2}$ or $\mu(GUT)=c M_{1/2}$ where $c$ is a proportionality constant. We obtain the value of parameter $c$ by fitting 
our points with a first degree polynomial curve with no constant, which is shown as a thin blue line. 
The value of parameter $c$ turns out to be 1.128, i.e., $\mu(GUT)=1.128 M_{1/2}$. Note that the value of parameter $c$ may depend on the distribution of points. Hence the points should be distributed as evenly as possible. We have tried our best to have evenly distributed points.  \\    
\\
$\bullet$ {\bf Part 2}: Now we calculate fine-tuning using Eq.~(\ref{fnpar}). Remember that here
the derivative $\frac{\partial Mz}{\partial M_{1/2}}$ is approximated by $\frac{\delta Mz}{\delta M_{1/2}}$.
For the justification of the approximation, the variation of $M_{1/2}$ and $M_{Z}$ should be very small compared to their original values.
We select the points for FT-calculations from evenly generated points in Part 1. We vary $\tan\beta$, $M_{1/2}$ and $M_{Z}$.
We insist on two requirements when we make the variation of $M_{1/2}$, $\tan\beta$ and $M_{Z}$:

\begin{enumerate}
\item The following GUT parameters vary as small as possible: $M_{GUT}$, 
 three Yukawa couplings $y_{i}(GUT)$ and three gauge couplings $g_{k}(GUT)$).
\item $\mu(GUT)$ generated in program is as close to the value  $\mu_{GUT} =1.128 M_{1/2}$ as possible.
\end{enumerate}
For the first requirement, we first analyze the variation of GUT parameters when $\tan\beta$, $M_{1/2}$ and $M_Z$ are varied and make sure that either they do not vary or the variation is negligible.
For the second requirement,  we set the criterion manually that the absolute value of relative change 
of $\mu(GUT)$, i.e., 
$|(\mu(GUT)^{'}-\mu(GUT))/\mu(GUT)|=|(\mu(GUT)^{'}-1.128 M_{1/2})/1.128 M_{1/2}|$ (where $\mu(GUT)^{'}$ is the value of $\mu(GUT)$ in any given run when we
vary $M_{1/2}$, $\tan\beta$ and $M_Z$ ) should be less than $0.00001$.
Thus, we vary $M_{1/2}$ and $M_Z$ and collected several hundred points which fulfill the above criterion. Maroon points in Fig.~\ref{ft} represent our results of fine-tuning measure calculations. We see that when the thick blue line (our data) and the thin blue line (the fitted line) are somewhat apart (at $M_{1/2}$ around 1000 GeV and 1800 GeV respectively), fine-tuning is also relatively large ($\sim$ 1.195 and 1.911 respectively). 
When these two lines meet, we see that fine-tuning measure is about 1 ($\Delta_{EENZ}\sim 1$). 



\begin{figure}[htp!]
\centering
\subfiguretopcaptrue
\subfigure{
\includegraphics[totalheight=5.5cm,width=7.cm]{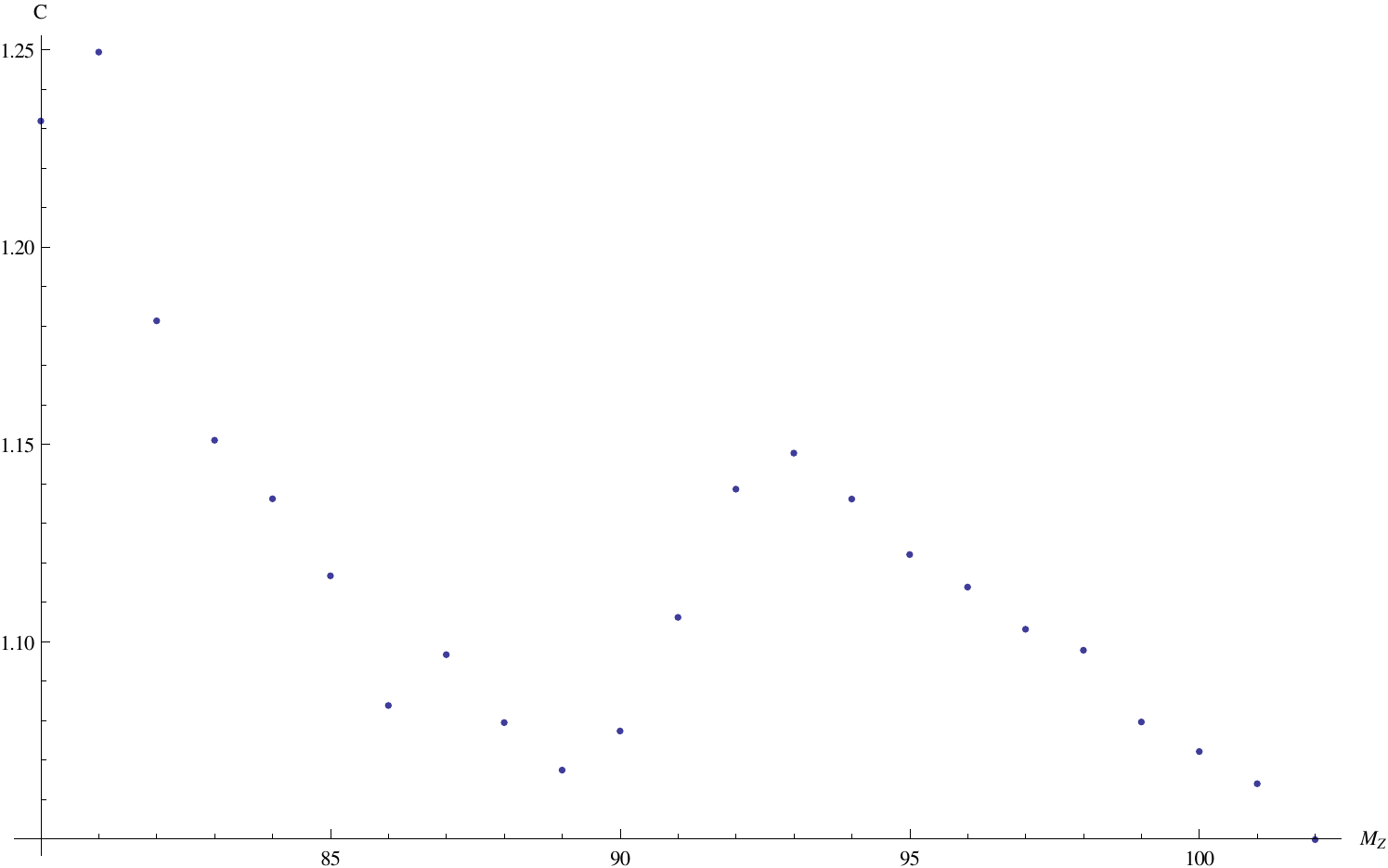}
}
\caption{
Parameter $c$ versus $M_Z$ plot.}
\label{c_mz}
\end{figure}

\begin{figure}[htp!]
\centering
\subfiguretopcaptrue
\subfigure{
\includegraphics[totalheight=5.5cm,width=7.cm]{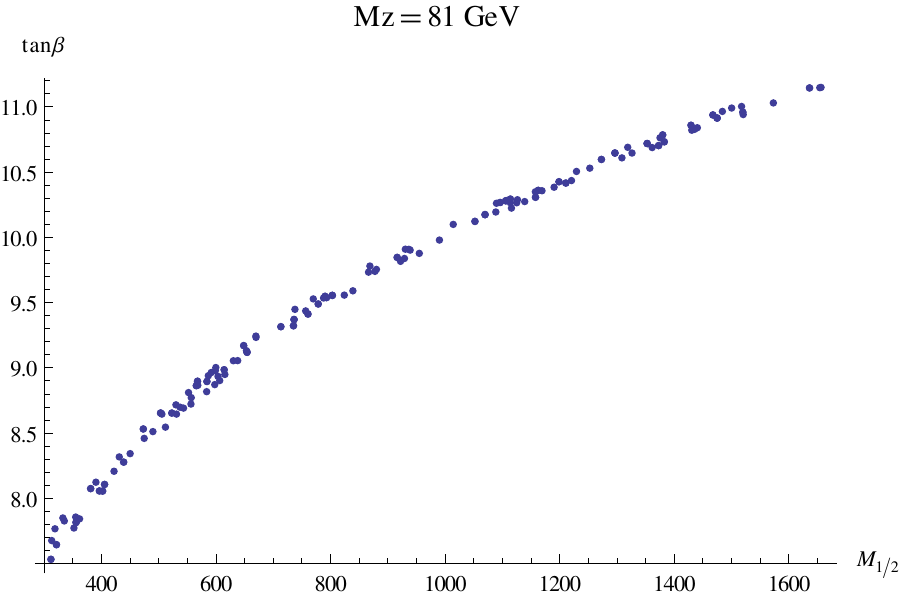}
}
\subfigure{
\includegraphics[totalheight=5.5cm,width=7.cm]{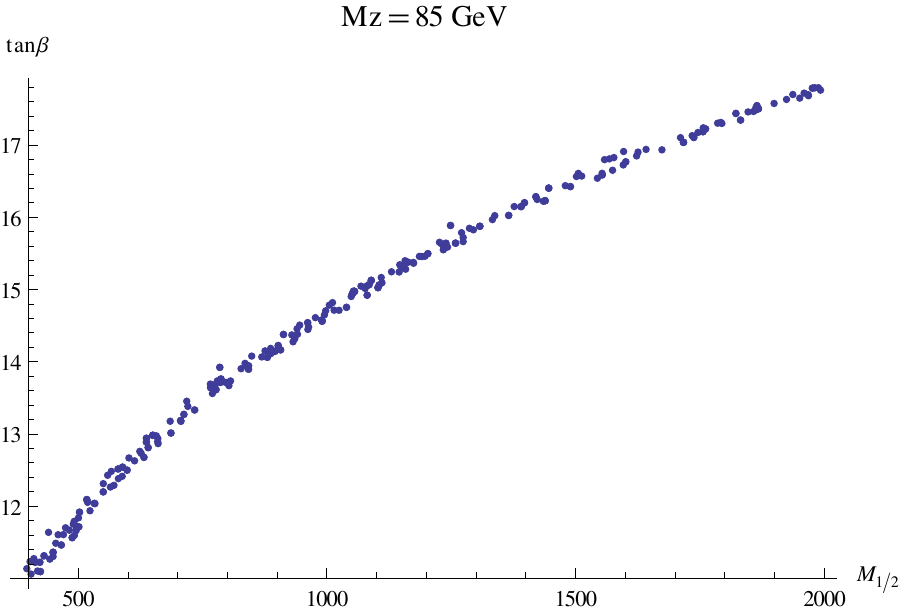}
}
\subfigure{
\includegraphics[totalheight=5.5cm,width=7.cm]{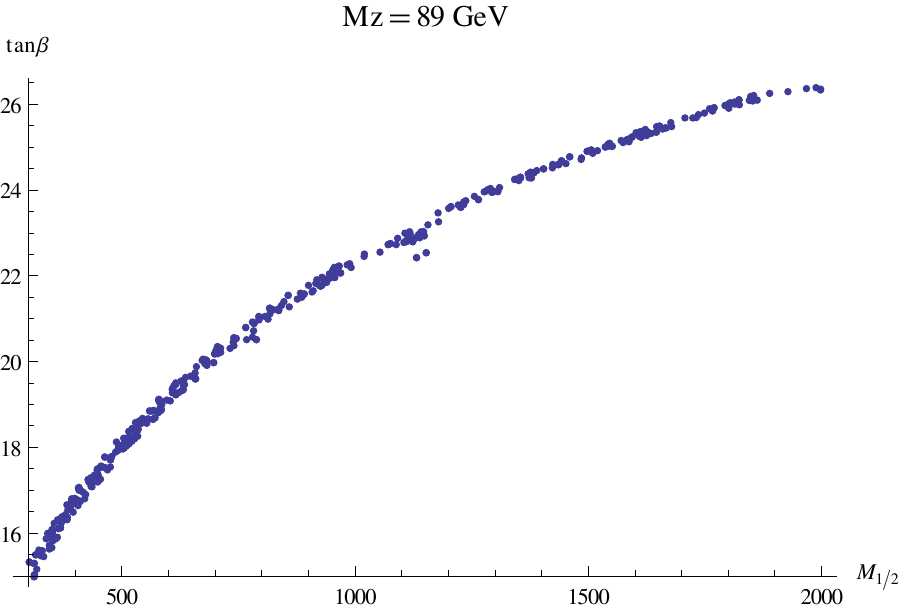}
}
\subfigure{
\includegraphics[totalheight=5.5cm,width=7.cm]{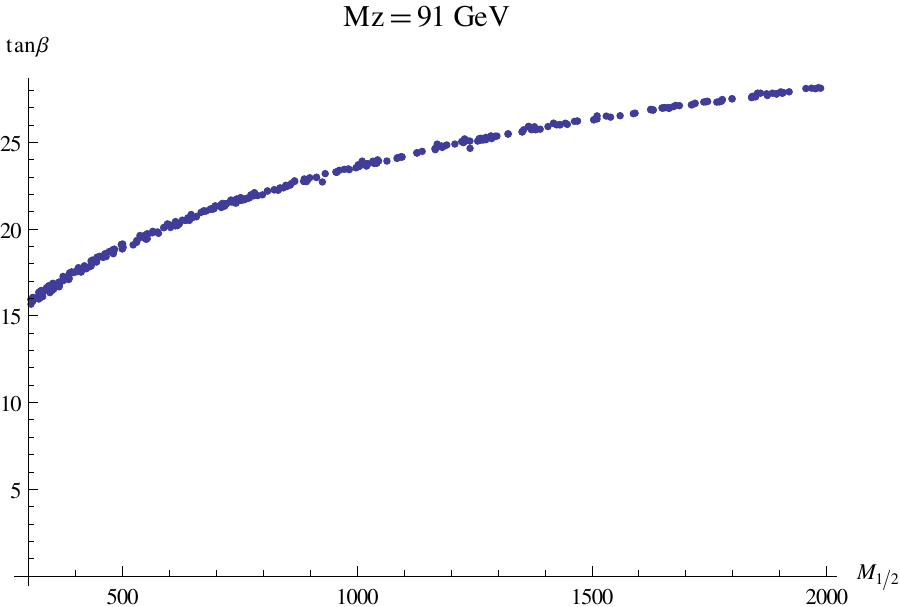}
}
\subfigure{
\includegraphics[totalheight=5.5cm,width=7.cm]{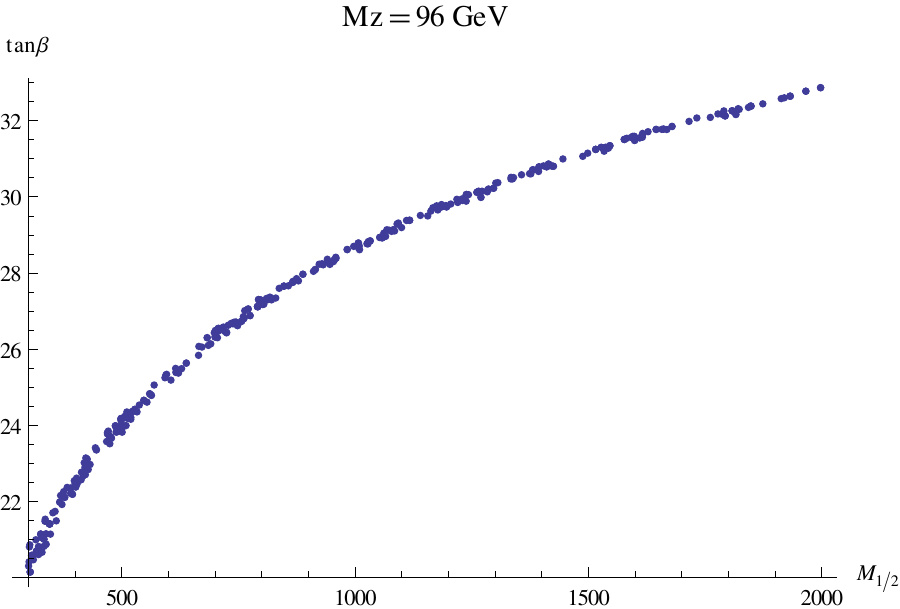}
}
\subfigure{
\includegraphics[totalheight=5.5cm,width=7.cm]{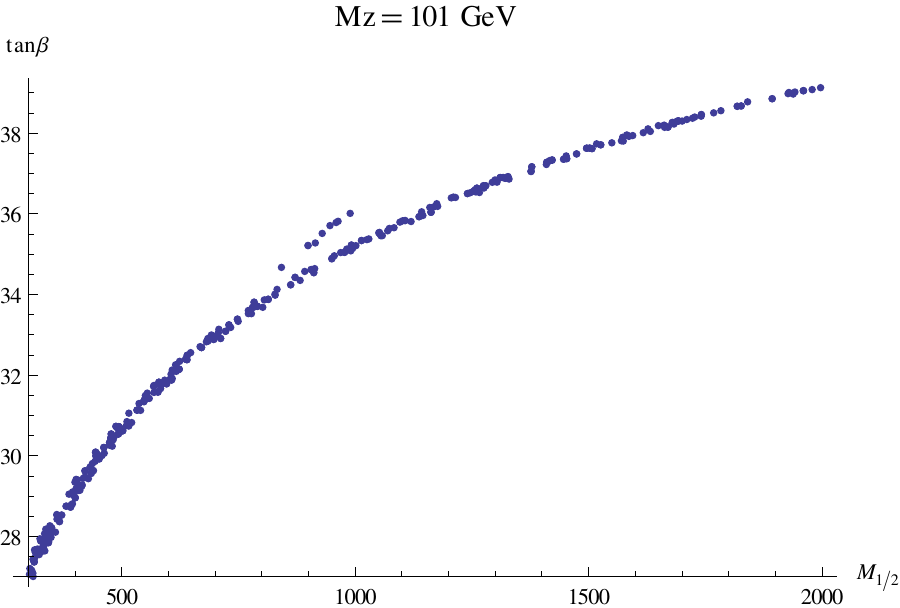}
}

\caption{
Plots in $\tan\beta$-$M_{1/2}$ plane. }
\label{tanb_m12}
\end{figure}

\begin{figure}[htp!]
\centering
\subfiguretopcaptrue
\subfigure{
\includegraphics[totalheight=5.5cm,width=7.cm]{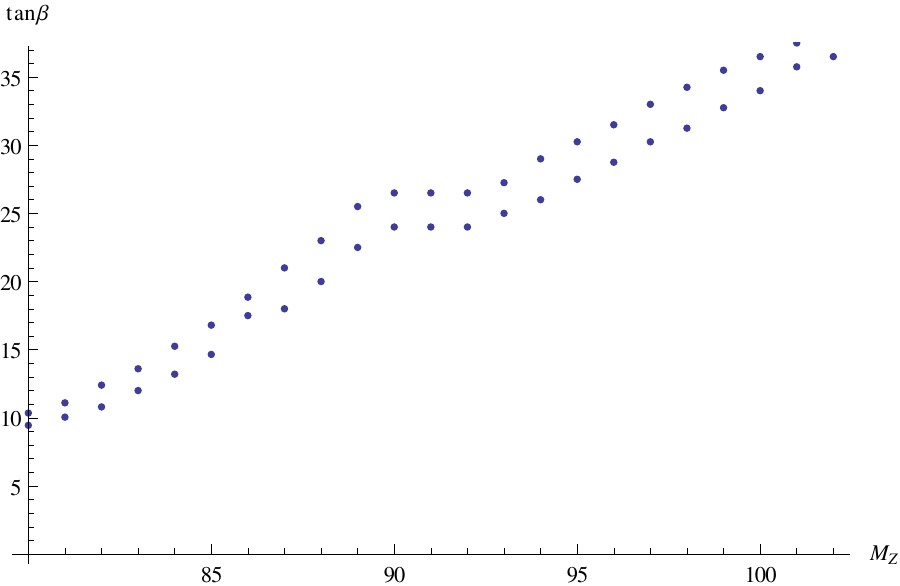}
}

\caption{
Plot in the $\tan\beta$-$M_Z$ plane.}
\label{tanb_mz}
\end{figure}

In addition, we study the dependence of parameter $c$ on the $Z$-boson mass $M_Z$ and present our results in 
Fig.~\ref{c_mz}. 
Here, we perform random scans by assuming $m_{0}=A_{0} = 0$ and Eq.~(\ref{input_param_range}) 
for individual values of $M_{Z}$ from 82 GeV to 102 GeV. Fig.~\ref{c_mz} shows various values of the parameter $c$ as a function of $M_Z$. We see that there is a couple of irregular points in the curve 
 at $M_Z$= 83 GeV and 86 GeV, respectively. In rest of the curve, we see that 
at $M_Z$= 82 GeV, the value of $c$ is around 1.23 and as the values of $M_{Z}$ increases, the value of $c$ decreases until $M_{Z}$= 89 GeV. After 
that $c$ rises with $M_{Z}$ for a short while until $M_{Z}$= 93 GeV and starts decreasing again as $M_{Z}$ increases to 102 GeV. In 
Fig.~\ref{tanb_m12}, we present graphs in $M_{1/2}-\tan\beta$ plane. Here we show that the initial values of $\tan\beta$ increases, though steadily, with $M_{1/2}$ for fixed $M_Z$, as well as with $M_Z$ implicitly. 
We display plots only for six values 
of $M_Z$ to show the trend, 
although we have generated data, as discussed above for individual values of $M_Z$ from 82 GeV to 102 GeV. 
Similarly, in Fig.~\ref{tanb_mz}, we present results in 
$M_Z$-$\tan\beta$ plane, where we select two different values of $M_{1/2}$, 800 GeV and 1600 GeV, and 
pick up the corresponding $\tan\beta$ values along with corresponding $M_Z$ values. This figure clearly 
shows the dependence of initial values of $\tan\beta$ on $M_Z$ values. The plateau that we see 
for $M_Z$ values 90 GeV to 93 GeV corresponds to the values of the parameter $c$ from 0.6 to 1.15.

\begin{figure}[htp!]
\centering
\subfiguretopcaptrue

\subfigure{
\includegraphics[totalheight=5.5cm,width=7.cm]{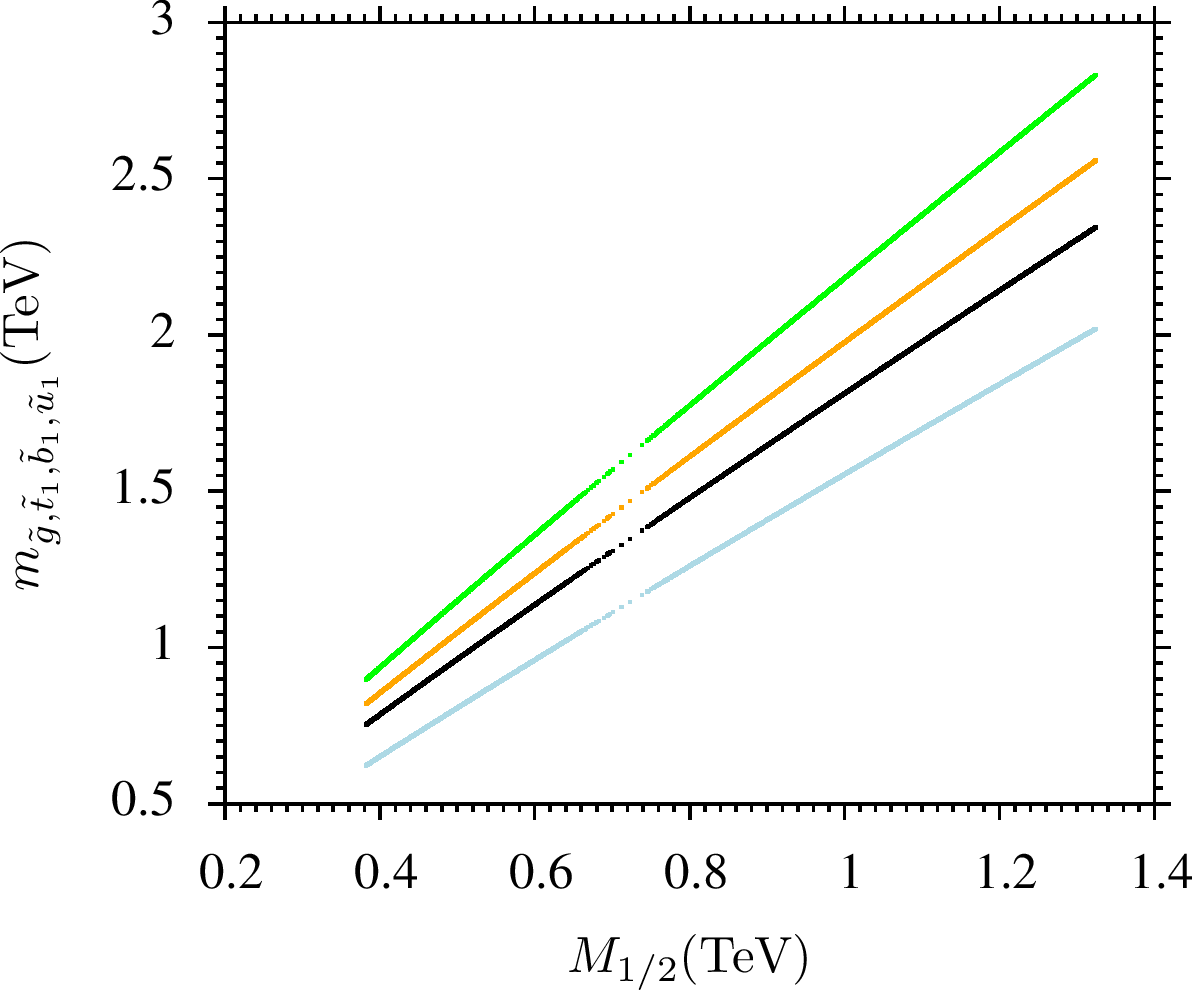}
}
\subfigure{
\includegraphics[totalheight=5.5cm,width=7.cm]{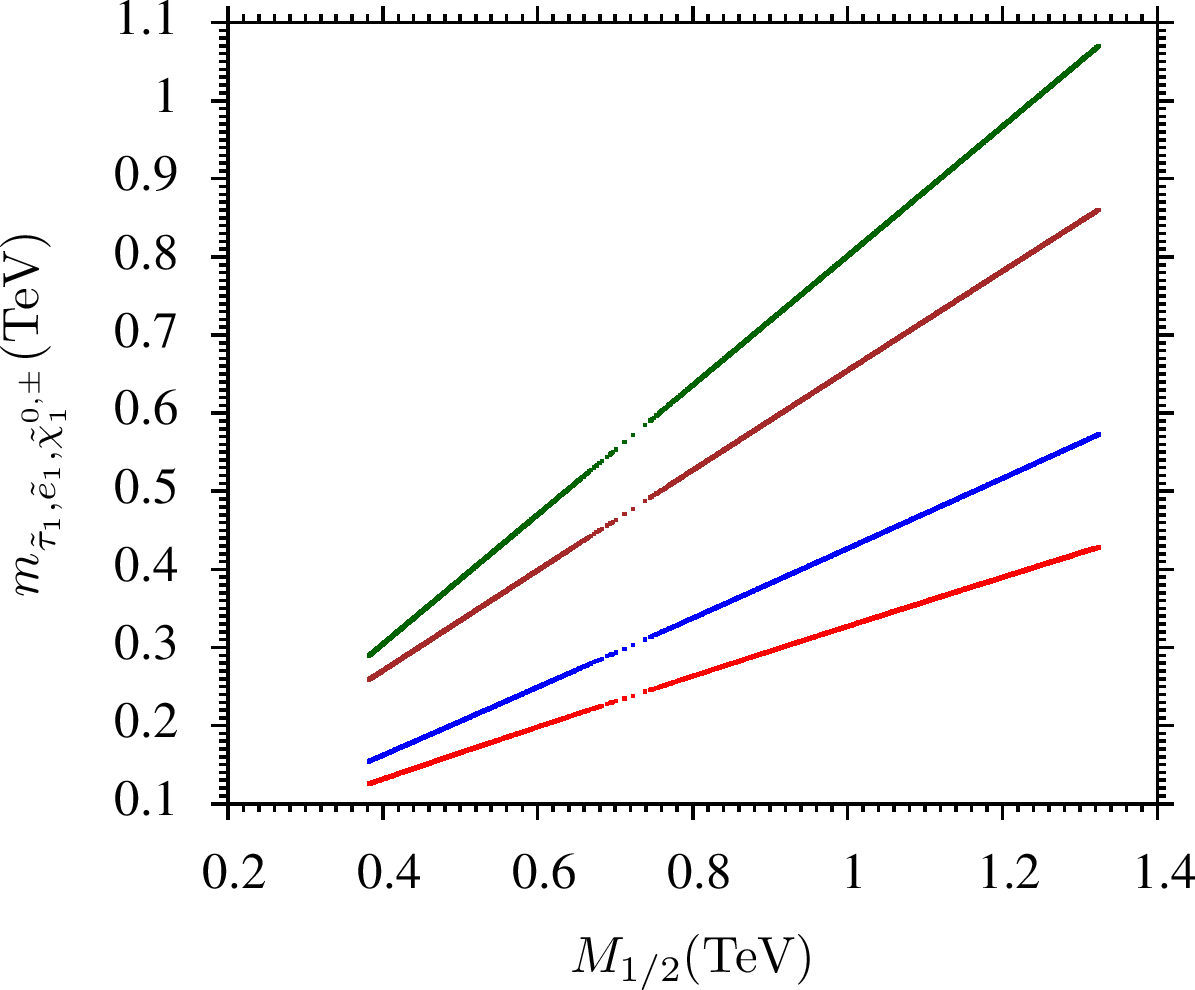}
}

\caption{Sparticle masses as functions of $M_{1/2}$. See text for color coding.
}
\label{spectrum}
\end{figure}

Moreover, we show sparticle masses as functions of $M_{1/2}$ in Fig.~\ref{spectrum}. 
In this figure, the value of parameter 
$c=\mu(GUT)/M_{1/2}$ varies only slightly in the range $c=1.128\pm0.001$. In the left panel 
we have a plot for color sparticles. The green, purple,
orange and black points represent $m_{\tilde g}$, $m_{\tilde t_1}$, $m_{\tilde t_1}$, $m_{\tilde u_1}$ and $m_{\tilde b_1}$, respectively. In the 
right panel, the dark green points represent $m_{\tilde \chi_{1}^{\pm}}$, blue points represent $m_{\tilde \chi_{1}^{0}}$, red points show 
$m_{\tilde \tau_{1}}$, and brown points depict $m_{\tilde e_{1}}$. Thus, all the sparticle masses are indeed 
linearly proportional to the gaugino mass.  We would like to emphasize that our calculations of $\Delta_{EENZ}$ for $n$=1 is equivalent to 
the traditional choice of $n$=2 since all the mass parameters are linearly proportional to $M_{1/2}$ in the No-Scale MSSM with the GM mechanism.
In fact, we generically have
\begin{eqnarray}
M_{Z}^n &=&f_n M_{1/2}^n~.
\end{eqnarray}
Therefore, $|\partial\ln(M_{Z}^n)/\partial\ln(M_{1/2}^n)|\simeq$ 1 is valid for any positive integer $n$.


\section{Dark Matter in No-Scale MSSM}
\label{sec:dm}

There are compelling evidences available in favor of the existence of dark matter (DM)~\cite{PDG}. 
Experiments suggest that it consists of electrically neutral stable 
massive particles~\cite{Ade:2013zuv,Hinshaw:2012aka}. In the MSSM with $R$-parity conservation, 
the lightest neutralino ($\tilde \chi_{1}^0$) can be the LSP and then the dark matter 
candidate~\cite{Goldberg:1983nd, Ellis:1983ew} 
(For reviews, see~\cite{Jungman:1995df, Bertone:2004pz}.). 
The observed DM relic density can be obtained if we assume that neutralino pairs annihilate 
through pseudo-scalar Higgs 
($A$-resonance), $Z$-boson ($Z$-resonance) or even through Higgs boson ($h$-resonance) 
to the SM particles in the early Universe. Also, the LSP neutralino 
can coannihilate with the NLSP, such as the lighter stau ($\tilde \tau_1$), lighter stop ($\tilde t_1$), 
chargino ($\tilde \chi_{1}^{\pm}$), etc, to reduce their number density and satisfy the observed 
relic density bounds. In the CMSSM/mSUGRA, even with these
mechanisms available to satisfy the relic density bounds, there is still a large parameter space where 
the relic density of neutralino is unusually large
 or the LSP is a charged sparticle like the lighter stau. In this scenario, to preserve the $R$-parity,
 one can assume 
that $\tilde \chi_{1}^0$ or $\tilde \tau_1$ are the NLSPs 
and can decay to the non-MSSM sparticles. For example, both the axino $\tilde a$, the fermionic partner of 
the axion and the gravitino, the fermionic partner of the graviton can be the LSP candidates.

In the no-scale MSSM, the LSP turns out to be the light stau, since
 the neutralino mass increases faster in RGE flow with $M_{1/2}$ as
compared to the lighter stau. In Ref.~\cite{Schmaltz:2000gy}, it was pointed out that the problem of a stau LSP can be
alleviated when the unification scale is raised sufficiently above the GUT scale, 
leaving the possibility open for a bino LSP as the dark matter 
candidate~\footnote{See~\cite{Ellis:2010jb} and references therein for phenomenological issues related 
to No-Scale models.}.

In our case, the gravitino should be very heavy as discussed above, 
so we are left with the $\tilde a$ LSP scenario. The axino is a promising dark matter candidate 
beyond the MSSM~\cite{Kim:1984yn, Bonometto:1993fx,Covi:1999ty,Covi:2001nw,Covi:2004rb,Brandenburg:2004du,Steffen:2008qp,Baer:2008yd}. It appears
when we extend the MSSM with the Peccei-Quinn mechanism~\cite{Peccei:1977hh,Peccei:1977ur} in order to 
solve the strong CP problem. The mass of axino
depends on the models and SUSY breaking scheme~\cite{Tamvakis:1982mw,Nieves:1985fq,Rajagopal:1990yx,Goto:1991gq}. 
The extremely weak interactions between the axino and charged slepton allow a long-lived NLSP 
charged slepton such as the lighter stau ($\tilde \tau_1$). These long-lived staus,
which were also called charged massive particles (CHAMPs) in Refs.~\cite{Drees:1990yw,Nisati:1997gb,Ambrosanio:1997rv,Feng:1997zr,Fairbairn:2006gg,Martin:1998vb}, can be probed at the LHC and have been studied extensively~\cite{LHCb:21 07 2014asa,Endo:2011uw,Ito:2010un,Ito:2010xj,Kitano:2010tt,Ito:2009xy,Pradler:2009mt,Jittoh:2009zzc,Yamanaka:2009zzb,Kohri:2008zza,Jittoh:2007fr,Ibarra:2006sz,Tarem:2005gxa}.

However, the Long-lived staus 
can cause problem and violate various cosmological bounds such as the Big Bang Nucleosyntheis bounds~\cite{Freitas:2009jb}, catalyst BBN bounds~\cite{Freitas:2009fb,pospelov,cyburt} and 
structure formation bounds~\cite{Polisensky:2010rw}. Thus, in such a scenario, one should have a 
mechanism through which the lifetime of the lighter stau via decay to the axino LSP is 
within reasonable time, and then the above mentioned cosmological bounds can be escaped. 
Such a scenario has been studied assuming stau as the NLSP and gravitino as the LSP in
Refs.~\cite{Benhenni:2011yt,Heisig:2013sva,Panotopoulos:2008cs,Feng:2004mt}. However,
the gravitino mass should be heavier compared to the gaugino mass
 in no-scale supergravity, and the parameter space satisfying 
all the constraints discussed above have heavy gaugino mass
$M_{1/2}\gtrsim$ 1800 GeV from the later study. Thus, we can only assume $\tilde a$ as the LSP.
 In the following, we  will explore the parameter space systematically 
where the light stau ($\tilde \tau_1$) is the NLSP 
and the axino ($\tilde a$) is the LSP with the no-scale SUGRA boundary conditions under various collider 
and cosmological constraints.

\subsection{Phenomenological Constraints}
\label{sec:scan}

Using the no-scale SUGRA boundary conditions, we perform random scans over the parameters space 
given in Eq.~(\ref{input_param_range}). 
Also, we consider $\mu > 0$, $m_t = 173.3\, {\rm GeV}$, and $m_b^{\overline{DR}}(M_{\rm Z})=2.83$ GeV. The initial data
are generated with SuSpect2.43~\cite{Djouadi:2002ze}. In this study we use the following parameter space:
\begin{align}
300\,\rm{GeV} \leq &   M_{1/2}   \leq 5000 \, \rm{GeV} ~,~\nonumber \\ 
1\leq & \tan\beta  \leq 60~.~
\label{input_param_range2}
\end{align}



The data points collected all satisfy the requirements of the radiative electroweak symmetry breaking (REWSB), no tachyonic sfermions or 
pseudo-scalar Higgs bosons.
We then employ MircOmegas3.6.9.2~\cite{micromegas} to calculate the relic density and other constraints.
After collecting the data, we require $|B_{0}|\lesssim$ 1 GeV and the following bounds 
on sparticle masses from the LEP2 experiment
\begin{eqnarray} 
m_{\tilde t_1},m_{\tilde b_1} \gtrsim 100 \; {\rm GeV} \\
m_{\tilde \tau_1} \gtrsim 105 \; {\rm GeV}  \\
m_{\tilde \chi_{1}^{\pm}} \gtrsim 103 \; {\rm GeV}
\end{eqnarray}
Also, we use Ref.~\cite{mamoudi}
 to implement the following B-physics constraints
\begin{eqnarray}
0.8\times 10^{-9} \leq{\rm BR}(B_s \rightarrow \mu^+ \mu^-) 
  \leq 6.2 \times10^{-9} \;(2\sigma)~~&\cite{Aaij:2012nna}& 
\\ 
2.99 \times 10^{-4} \leq 
  {\rm BR}(b \rightarrow s \gamma) 
  \leq 3.87 \times 10^{-4} \; (2\sigma)~~&\cite{Amhis:2012bh}&  
\\
0.15 \leq \frac{
 {\rm BR}(B_u\rightarrow\tau \nu_{\tau})_{\rm MSSM}}
 {{\rm BR}(B_u\rightarrow \tau \nu_{\tau})_{\rm SM}}
        \leq 2.41 \; (3\sigma)~~&\cite{Asner:2010qj}&  
\end {eqnarray}
In addition, we impose the following bounds from the LHC as well
\begin{eqnarray}
m_h  = 123-127~{\rm GeV}~~&\cite{ATLAS, CMS}&  \\ 
m_{\tilde{g}}\gtrsim 1.7 \, {\rm TeV}\ ({\rm for}\ m_{\tilde{g}}\sim m_{\tilde{q}}) &
\cite{Aad:2014wea,Chatrchyan:2013wxa}\\
m_{\tilde{g}}\gtrsim 1.3 \, {\rm TeV}\ ({\rm for}\ m_{\tilde{g}}\ll m_{\tilde{q}}) &
\cite{Aad:2014wea,Chatrchyan:2013wxa}\\
\end{eqnarray}
As far as the muon anomalous magnetic moment $a_{\mu}$ is concerned, we require that the benchmark
points be at least as consistent with the data as the Standard Model.
\subsection{Numerical Results}
\label{results}

\begin{figure}[htp!]
\centering
\subfiguretopcaptrue

\subfigure{
\includegraphics[totalheight=5.5cm,width=7.cm]{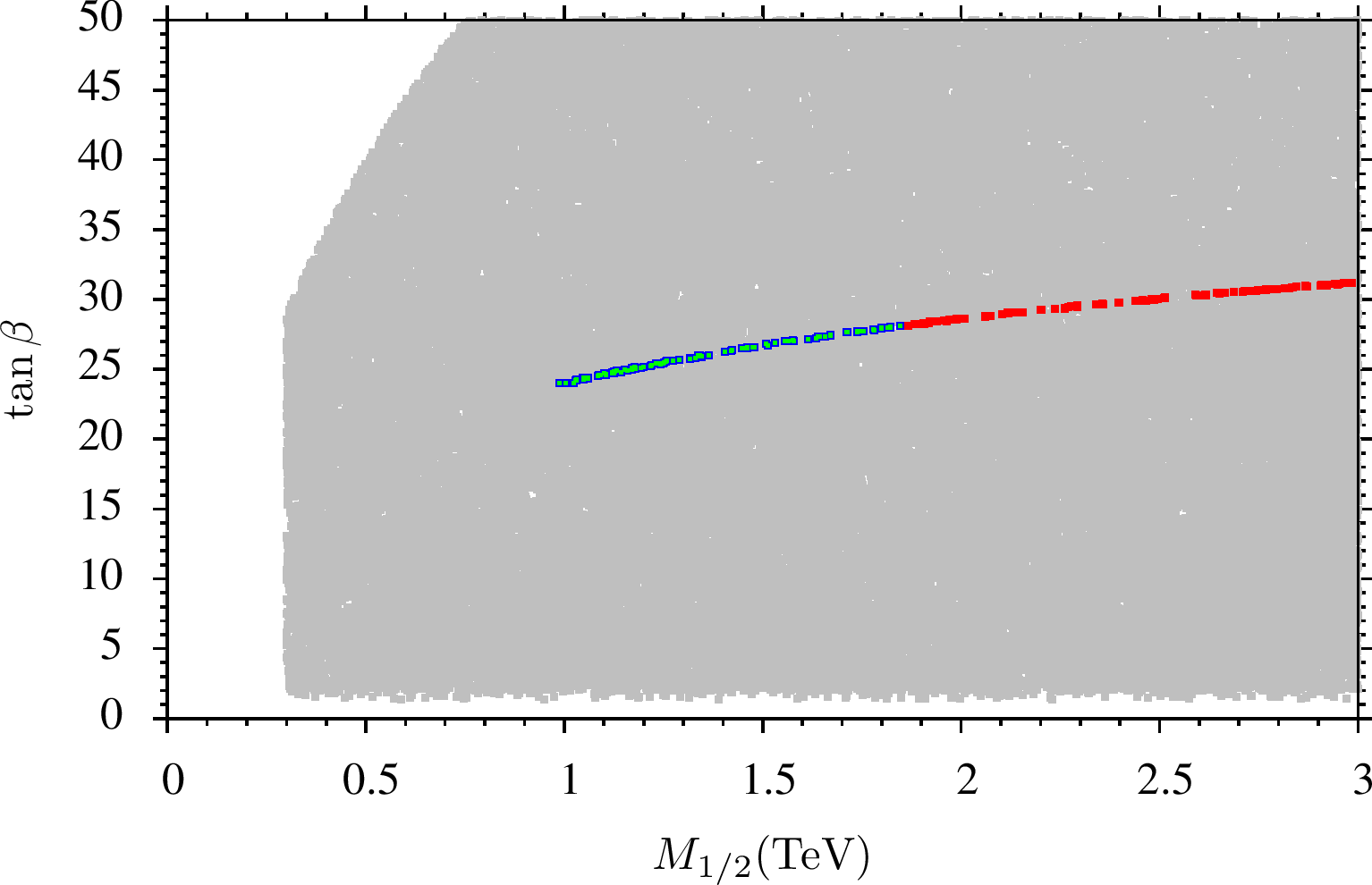}
}
\subfigure{
\includegraphics[totalheight=5.5cm,width=7.cm]{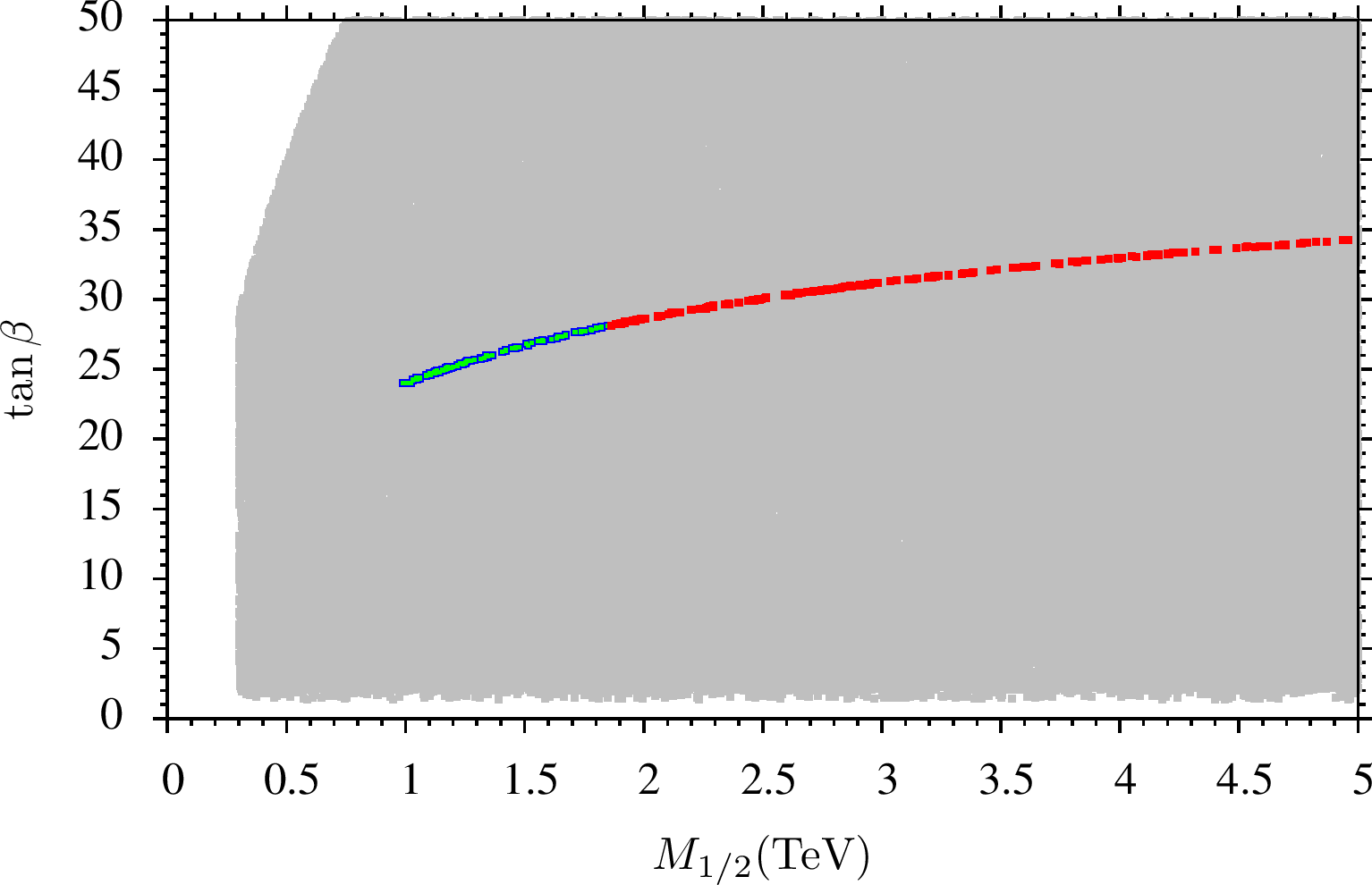}
}
\caption{
The viable parameter spaces in $M_{1/2}-\tan\beta$ plane.
The gray points have the successful REWSB and no tachyonic sfermions or pseudo-scalar Higgs bosons. 
The blue points further satisfy sparticle mass bounds, 
green points form a subset of blue points and satisfy B-physics constraints, and red points 
are subset of green points and satisfy Higgs mass bounds.
}
\label{m12tanb}
\end{figure}

Here we present results of our scans. In the left panel of 
Fig.~\ref{m12tanb}, we show plot in $M_{1/2}$-$\tan\beta$ plane for $M_{1/2} \lesssim 3000$~GeV. 
The gray points satisfy the constraints of the REWSB and no tachyonic sfermions or pseudo-scalar Higgs bosons, 
blue points satisfy the sparticle mass bounds, green points form a subset of blue points and satisfy 
B-physics bounds, and red points form a subset of 
green points and are consistent with $m_{h}=125 \pm 2$ GeV constraint. In this plot we see that in the left top corner around 
$\tan\beta \gtrsim$ 30, we do not have 
points with successful REWSB.  We note that the application of the constraint $|B_{0}|\lesssim$ 1 GeV reduces the parameter space to a strip of 
points like Fig.~\ref{tanb_m12}. We also notice that a sharp cut in blue points around $M_{1/2}\sim$ 1 TeV
 due to the requirement $m_{\tilde q}\gtrsim$ 1800 GeV. When we apply the Higgs mass bounds (red points),
in our present data we need $M_{1/2}\gtrsim$ 1800 GeV and $\tan\beta\gtrsim$ 27. As we discussed earlier, 
in no-scale MSSM, the gravitino would be very heavy due to $M_{3/2}\propto M_{1/2}$. 
This clearly shows why we have chosen axino LSP scenario instead of gravitino LSP scenario. 
For comparison, in the right panel, we also 
display the same graph with $M_{1/2} \lesssim$ 5000~GeV. We see in this figure that we have more red points 
with slight increase in $\tan\beta$ range.
For our further analyses, we will restrict ourselves with the data $M_{1/2}\lesssim$ 3000 GeV. 
As we will show in the benchmark point Table~\ref{table1} that
even with $M_{1/2}\lesssim$ 3000 GeV, we already have heavy spartcile spectra.

\begin{figure}[htp!]
\centering
\subfiguretopcaptrue

\subfigure{
\includegraphics[totalheight=5.5cm,width=7.cm]{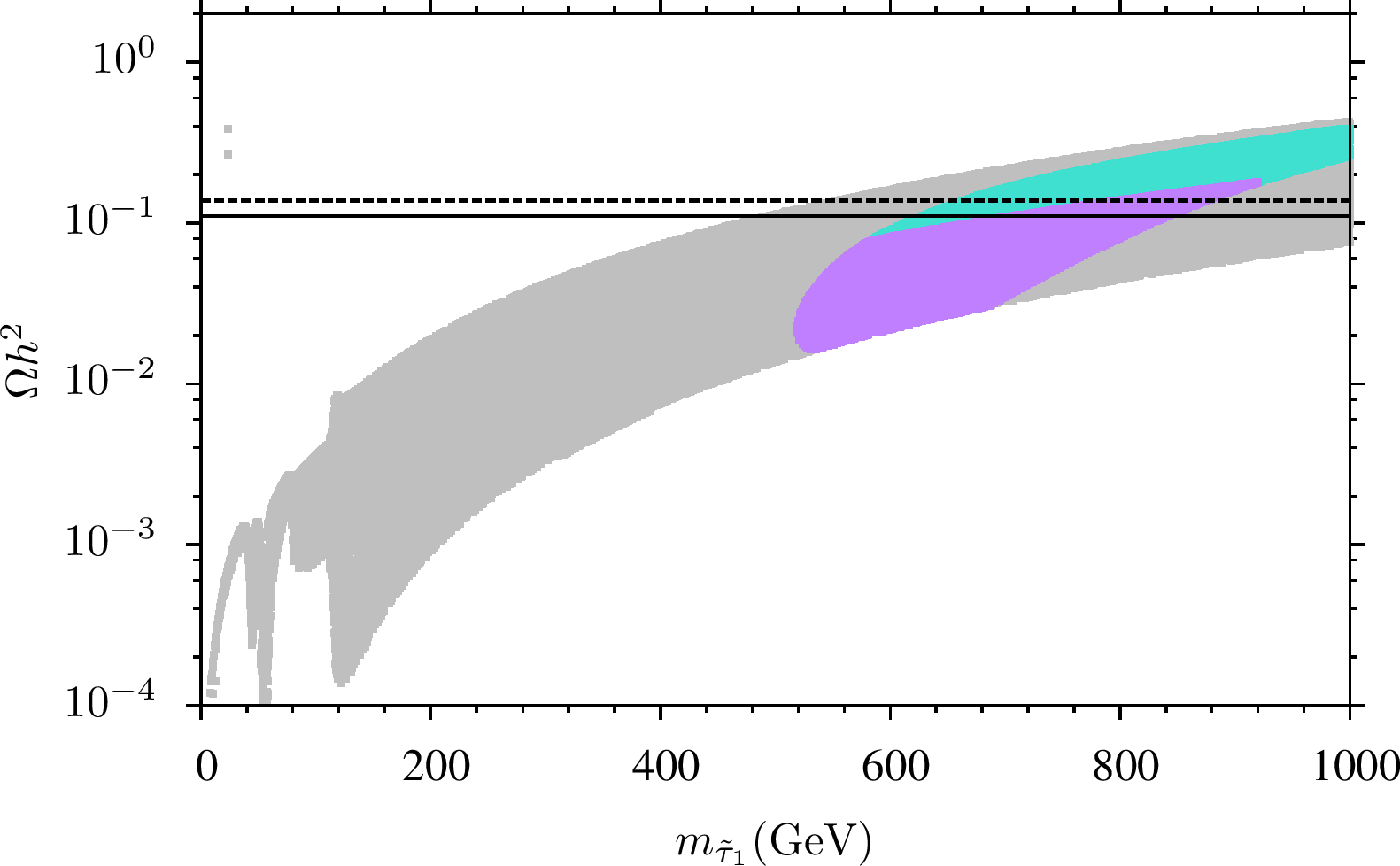}
}
\caption{
 $m_{\tilde \tau_1}$ versus $\Omega h^{2}$. 
The gray points have the successful REWSB and no tachyonic sfermions or pseudo-scalar Higgs bosons. 
The aqua  points further satisfy the sparticle mass bounds, as well as the
B-physics and Higgs mass bounds. And the purple points are subset of
 aqua points and satisfy $|B_{0}|\lesssim$ 1 GeV.
The horizontal black solid line represents $\Omega h^{2}$=0.11, while dashed black line shows $\Omega h^{2}$=0.137.}
\label{oh2}
\end{figure}

\begin{figure}[htp!]
\centering
\subfiguretopcaptrue

\subfigure{
\includegraphics[totalheight=5.5cm,width=7.cm]{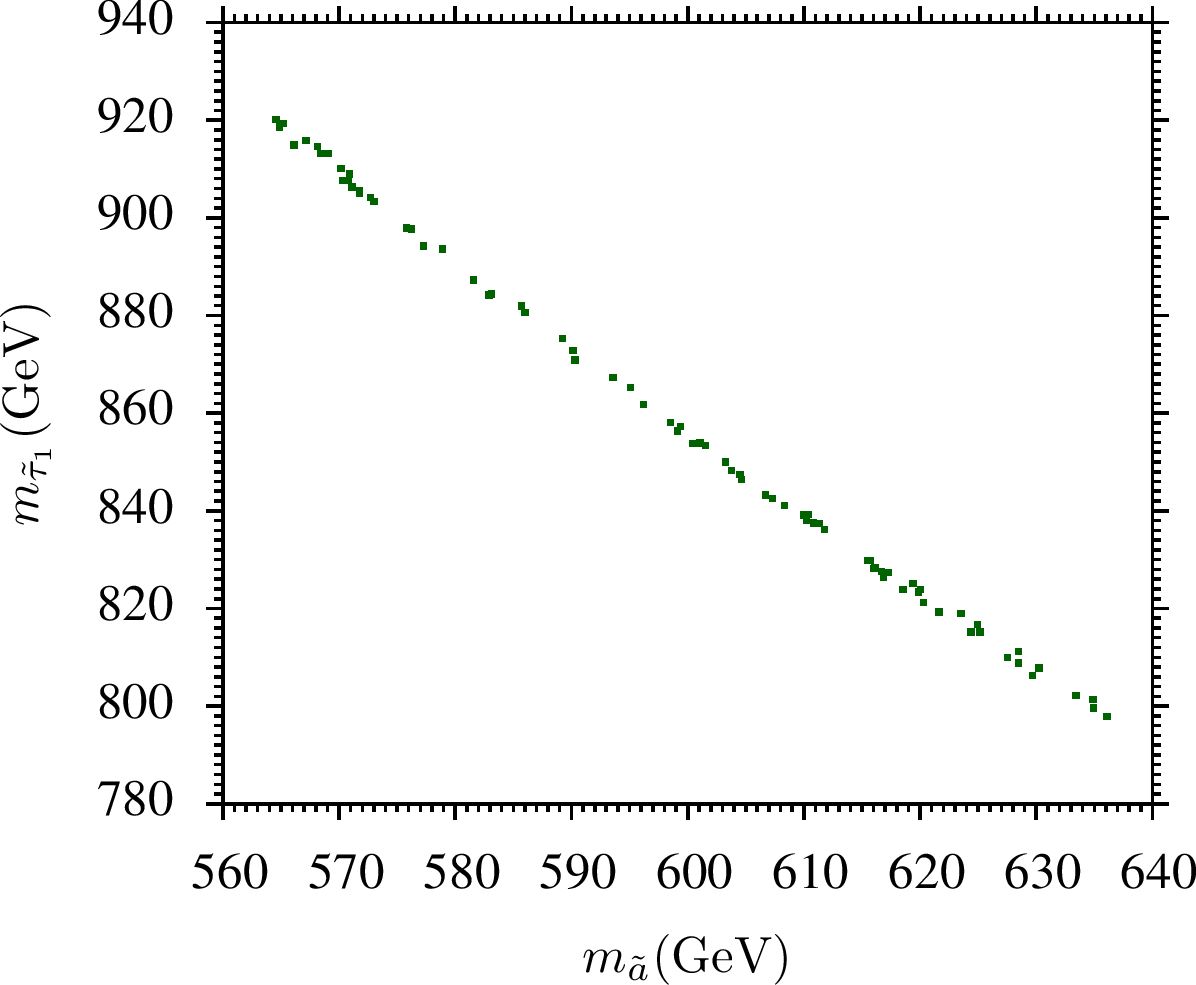}
}

\caption{
$m_{\tilde a}$ versus $m_{\tilde \tau_{1}}$.}
\label{axino_stau}
\end{figure}

\begin{figure}[htp!]
\centering
\subfiguretopcaptrue

\subfigure{
\includegraphics[totalheight=5.5cm,width=7.cm]{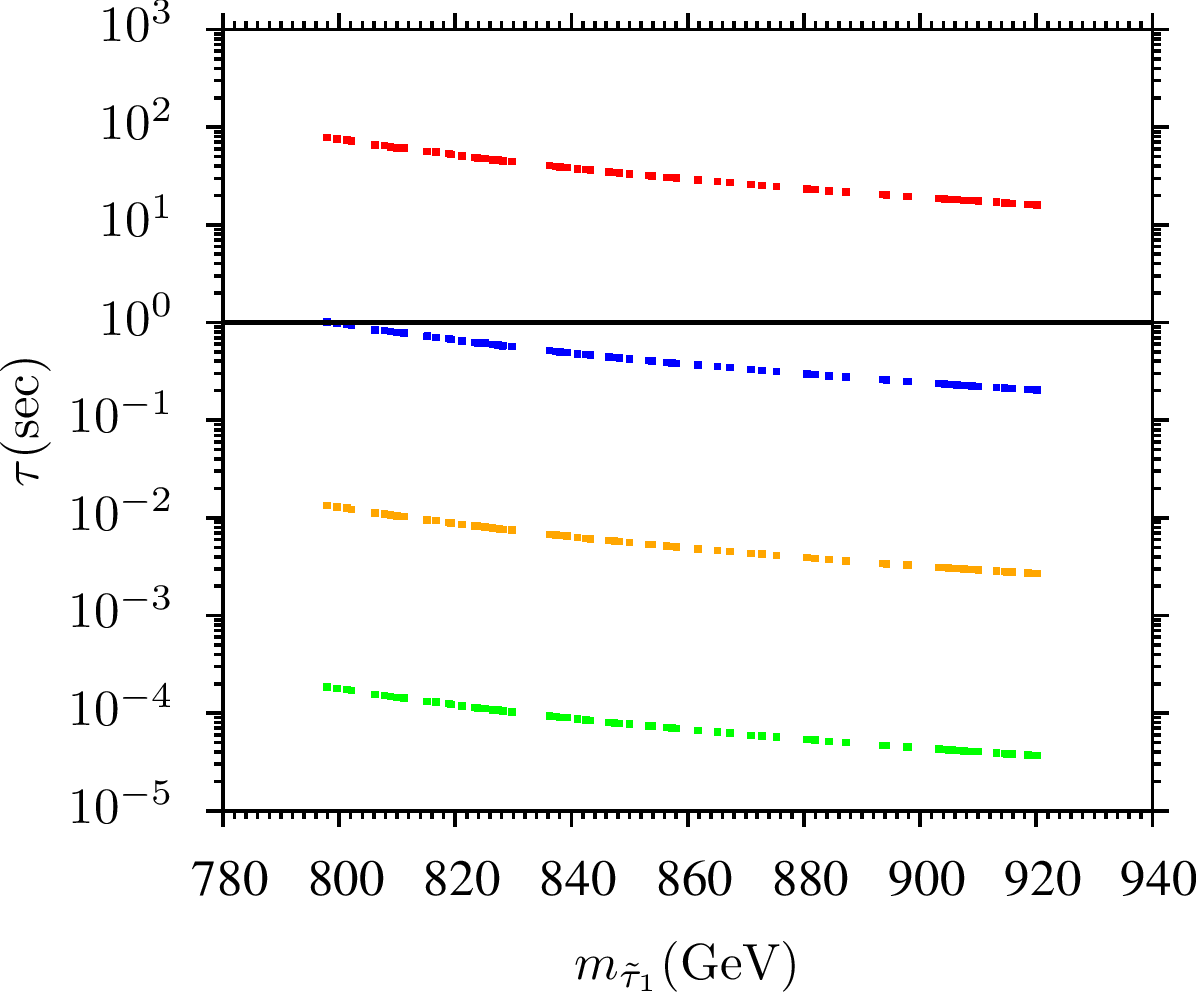}
}

\caption{
The lifetimes of staus ($\tau_{\tilde \tau_1}$) in seconds ($s$) as functions of stau masses $m_{\tilde \tau_1}$ 
for $|e_Q|$=1/3 and $y$=1.
Here, the light-green, orange, blue and red points represent the axion decay constants
$f_a=10^{9}, ~10^{10}, ~10^{11}$,  and $10^{12}$ GeV, respectively.
The horizontal black solid line represents $\tau_{\tilde \tau_1}$= 1 second.}
\label{lifetime}
\end{figure}

\begin{figure}[htp!]
\centering
\subfiguretopcaptrue

\subfigure{
\includegraphics[totalheight=5.5cm,width=7.cm]{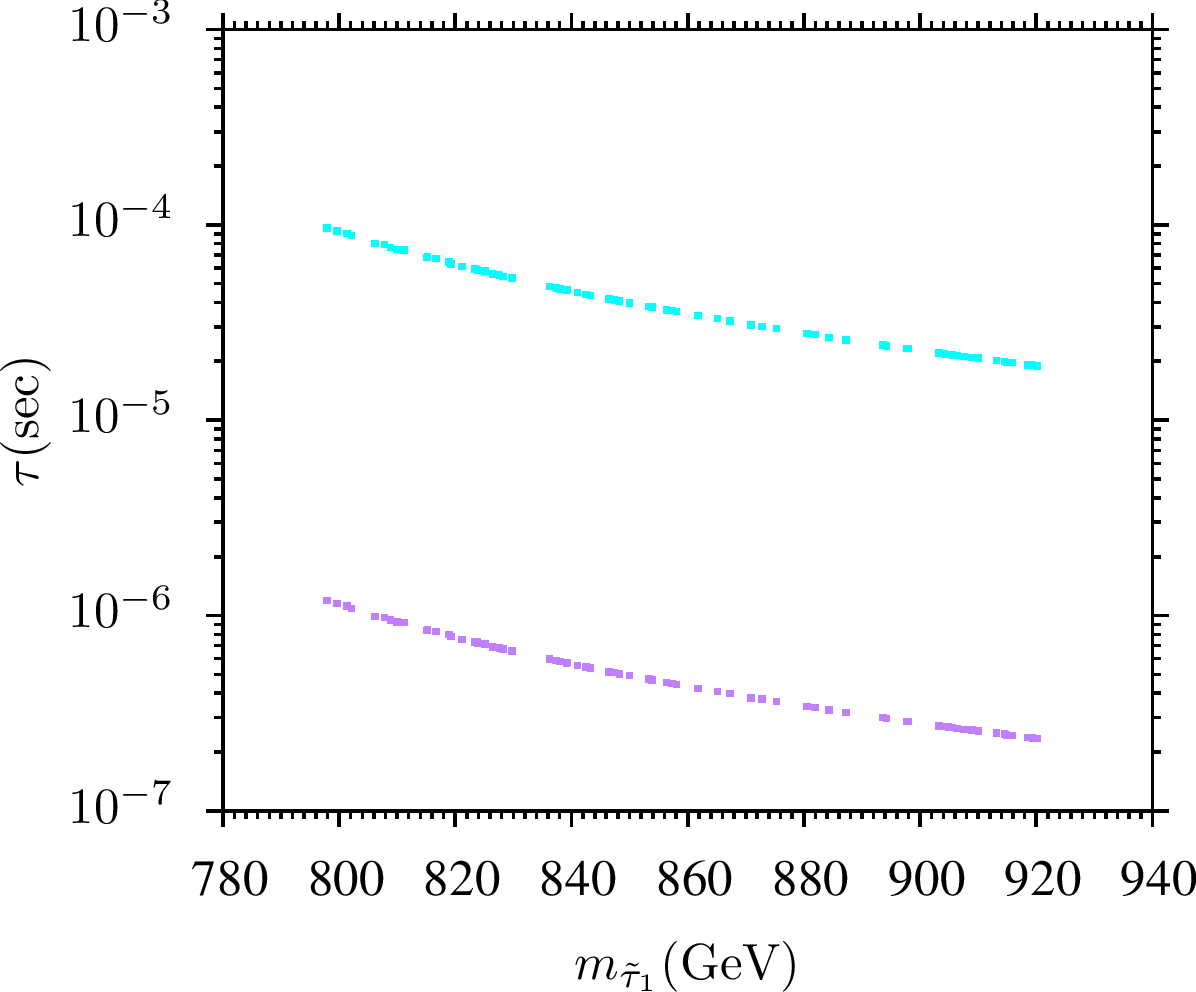}
}
\subfigure{
\includegraphics[totalheight=5.5cm,width=7.cm]{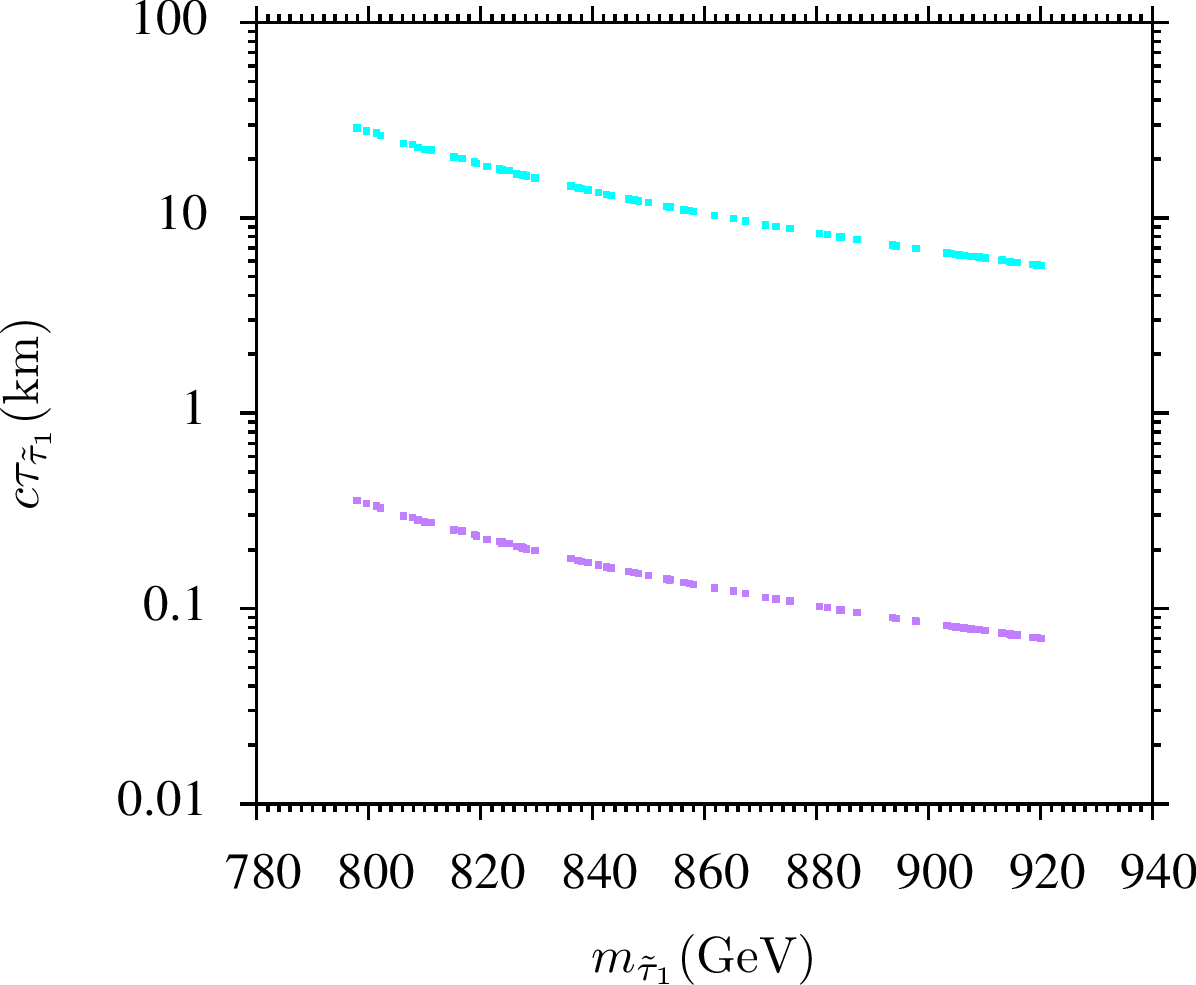}
}
\caption{
In the left panel, the lifetimes of staus ($\tau_{\tilde \tau_1}$) in seconds as functions of stau masses $m_{\tilde \tau_1}$ 
for $y$=1 and $f_{a}=7 \times 10^{8}$ GeV. The aqua and purple points respectively represents $|e_Q|=1/3$ and $|e_Q|=1$.
In the right panel, with the same color coding, corresponding decay lengths ($c\tau_{\tilde \tau_{1}}$) 
in kilometers (km) as  functions of $m_{\tilde \tau_1}$.
\label{fa8}}
\end{figure}

For our DM studies we assume non-thermally produced (NTP) axino LSP from decays of stau NLSP. 
We assume that each stau NSLP decays to an axino. The relic density of such axinos satisfy 
the observed DM relic density, $\Omega_{\tilde a}^{NTP}\simeq \Omega_{DM}$. As we mentioned earlier, 
because of extremely weak interaction with axino LSP, stau NLSP can be
long-lived. Since the long-lived charged particles like stau can disturb successful predictions 
of the BBN and violate the other cosmological bounds, it is 
essential to calculate the lifetime of stau NLSP. This requires the knowledge of
the axino mass. Usually the axino mass is a model dependent 
parameter. Note that axino is not a particle in the MSSM, 
we calculate its mass in the following way. As we just stated, we assume the observed dark matter relic density
$\Omega_{\tilde a}^{NTP}=0.11$~\cite{Hinshaw:2012aka} 
and use the following relation to compute the axino mass
\begin{equation}
m_{\tilde a}=m_{NLSP}\frac{\Omega_{\tilde a}^{NTP} h^{2}}{\Omega_{NLSP} h^{2}}~.~\,
\label{eqoh2}
\end{equation}  
In order to have $m_{NLSP}$ (in our case the light stau is the NLSP) and $\Omega_{NLSP} h^{2}$, 
we present the results in $m_{\tilde \tau_1}-\Omega_{NLSP} h^{2}$ plane in Fig.~\ref{oh2}.
The gray points represent successful REWSB and no tachyonic sfermions or pseudo-scalar Higgs bosons. 
The aqua  points satisfy sparticle mass bounds, B-physics and Higgs mass bounds. 
The purple points are subset of aqua points and satisfy
$|B_{0}| \lesssim$ 1 GeV constraint. Black solid line represent observed dark matter relic density while black dashed line show the upper WMAP9 5$\sigma$ bound.
We pick up those points which have $\Omega_{NLSP} h^{2}$ greater than the WMAP9 5$\sigma$ bounds. 
This gives us both the lighter stau mass and the value of 
relic density to be used in Eq.~(\ref{eqoh2}). We show the NLSP stau mass and corresponding  LSP axino mass 
in Fig.~\ref{axino_stau}.  Here, 
we see that the axino LSP mass range [565,~638] GeV corresponds to the stau NLSP mass range [788,~920] GeV. The stau NLSP (in our case 
the right-handed stau $\tilde \tau_R$)  can decay into the axino via $\tilde \tau_R \rightarrow \tau \tilde a$ (2-body decay), 
$\tilde \tau_R \rightarrow \tau \tilde a \gamma$ (3-body decay) and $\tilde \tau_R \rightarrow \tau \tilde a q {\bar q}$ (4-body decay). 
The lifetime of stau NSLP is governed by the 2-body decay, while the contribution
of 3-body decay remains below about 3$\%$. The 4-body decay channels like $\tilde \tau_R \rightarrow \tau \tilde a q {\bar q}$, 
$\tilde \tau_R \rightarrow \tau \tilde a \gamma \gamma$ and $\tilde \tau_R \rightarrow \tau \tilde a l^{+} l^{-}$ have negligible impact on the total 
stau decay rate as they are suppressed by an additional factor of $\alpha$ compared to the 3-body decay. This is why we do not consider them in our present study. The calculations of the decay rates $\Gamma (\tilde \tau_R \rightarrow \tau \tilde a)$ and 
$\Gamma(\tilde \tau_R \rightarrow \tau \tilde a \gamma)$ have been reported in \cite{Brandenburg:2005he,Freitas:2011fx} considering SUSY hadronic or 
KSVZ axion models~\cite{Kim:1979if,Shifman:1979if}. In \cite{Brandenburg:2005he}, these calculations were done using an effective 
theory where (s)quark were integrated out, and then may lead to logarithmic divergences. It was also shown that these logarithmic divergences can be regulated 
with the cutoff $f_a$, and only the dominant contributions are kept. In addition, a factor $\xi$ and a mass scale $m$ 
are introduced to parametrize
the uncertainty associated with this cutoff procedure. We will use the decay rates given in~\cite{Freitas:2011fx} in our work because of absence of such 
issues. In the 
calculation of $\Gamma (\tilde \tau_R \rightarrow \tau \tilde a)$, the scalar
one-loop self-energy function ($B_0)$ and vertex function ($C_0$)~\cite{'tHooft:1978xw} are  evaluated by
the LoopTools package~\cite{Hahn:1998yk}.
For the calculations of
$\Gamma (\tilde \tau_R \rightarrow \tau \tilde a)$ we use Eqs.~(3.1-3.3) and for $\Gamma(\tilde \tau_R \rightarrow \tau \tilde a \gamma)$ we use Eqs.~(3.4-3.6) in Ref.~\cite{Freitas:2011fx}. In numerical calculations we take $\alpha=\alpha^{\overline{MS}}(m_Z)=1/128$, $\sin^{2}\theta_W =1-m^2_{W}/m^2_{Z}=0.2221$, $|e_{Q}|=1/3$ and $y$= 1, where $|e_{Q}|$ and $y$ are the electric charge and Yukawa coupling of the new heavy KSVZ quarks~\cite{Freitas:2011fx}. 
The lifetime ($\tau_{\tilde \tau_1}$) of the light stau as a function of its mass $m_{\tilde \tau_1}$ for four different values of $f_a$ is shown in Fig.~\ref{lifetime}. The green, orange, blue and red points represent solutions 
for  the axion decay constants
$f_a=10^{9}, ~10^{10}, ~10^{11}$,  and $10^{12}$ GeV, respectively. The solid
black line represents $\tau_{\tilde \tau_1}$ =1 second as a reference line. As expected, we see that the 
lifetimes increase with $f_a$ values. For 
$f_a=10^{9}$ GeV, $10^{11}$ GeV, and $10^{12}$ GeV, the lifetime of light stau is
 less than about $10^{-4}$ second, 1 second, 100 seconds, respectively.
In Fig.~14 of~\cite{Freitas:2011fx}, plots in $m_{\tilde a}-m_{\tilde \tau_1}$ plane are 
shown with $f_a=10^{12}, ~3\times 10^{12}, ~10^{13}$ GeV, $|e_Q|$= 1/3 and
$m_{bino}=1.1~m_{\tilde \tau_1},1.01~m_{\tilde \tau_1}$. The authors of Ref.~\cite{Freitas:2011fx} have given 
a very comprehensive presentation of various cosmological constraints on stau NLSP and axino LSP masses. 
In Fig.~14 of~\cite{Freitas:2011fx}, they have
considered $\Omega_{\tilde a}^{NTP}=\Omega_{DM}$ and implemented structure formations bounds, primordial nucleosynthesis bounds (both the hadronic and electromagnetic bounds). In order to compare our results with Fig.~14 of 
Ref.~\cite{Freitas:2011fx}, we have to keep in view both Figs. \ref{axino_stau} and \ref{lifetime} 
of our paper. First
of all we note that our solutions lie in two narrow windows $m_{\tilde a}\sim$ [565,~638] GeV 
and $m_{\tilde \tau_1}\sim$ [788,~920 GeV]. From 
Fig.~14 of~\cite{Freitas:2011fx}, we see that the structure formation bounds exclude axino mass 
$m_{\tilde a} \lesssim$ 100 GeV for $f_{a} \sim 10^{12}$ GeV
and axino mass around 250 GeV for $f_a=10^{13}$ GeV. Moreover, the hadronic BBN constraints (both the conservative and severe) imposed by the deuterium (D) abundance mostly remain important for axino mass less than 100 GeV but rise up to higher axino mass ranges $\sim$ 500 GeV for $f_a=10^{13}$ GeV. Similarly,  the corresponding electromagnetic BBN constraint 
remains important for stau mass less than 400 GeV where axino mass is in the range 1 GeV
 to 300 GeV for $f_a$ values $10^{12}$ and $10^{13}$ GeV. In scenario where stau NLSP has the lifetime greater than $10^{3}$ s, negatively charged stau
can form bound states with the primordial nuclei and thereby catalyse the formation of ${^6}$Li and ${^9}$Be in the early Universe~\cite{pospelov}, which is
why they are called catalyse BBN (CBBN). From Fig.~14 of Ref.~\cite{Freitas:2011fx}, we see that our stau NLSP having lifetime $\gtrsim 10^{3}$ s may be ruled out.
As far as the stau NSLPs with lifetime $\tau_{\tilde\tau_1} \lesssim$ 100 s is concerned, as also indicated in Ref.~\cite{Freitas:2011fx}, one can expect the
mild constraints as compared to above mention bounds. We have not found the detailed study discussing 
the cosmological constraints for 
$\tau_{\tilde\tau_1} \lesssim$ 100 s. When we compare our results with the above
observations, we note that our solutions seem fine although there is a slight difference in assumption 
(we do not have fixed the ratio between bino mass and 
stau mass). Maybe a separate analysis is needed to study the stau NLSP in no-scale MSSM with lifetime greater 
than 100 s with updated astrophysical 
constraints. Note that 
Fig.~\ref{lifetime} shows that $\tau_{\tilde\tau_1}$ depends on the $m_{\tilde \tau_1}$, $f_a$ and $|e_Q|$. 
From above discussions, it is evident that the cosmological bounds indeed restrict the parameter spaces 
of the SSMs with stau NLSP and axino LSP.  Moreover, one can probe and study the properties of 
such axinos at colliders. As mentioned earlier, since axino interactions are extremely weak, the direct 
detection of axinos seems hopeless. Likewise, their direct production at colliders is very strongly suppressed. 
Instead, one expects a large sample of the
NLSPs from the pair productions or cascade decays of heavy sparticles, provided the NLSP belongs to the MSSM spectrum.
In our case, the stau NLSPs can be produced in cascade decays of heavy sparticles or from directly pair produced. 
One can measure the mass of staus using 
time-of-flight data from the muon chambers.
In addition, it is essential to study the decays of staus to measure
 $\tau_{\tilde\tau_1}$ and axino mass. Our analysis given above clearly shows that
the lifetime of stau NLSP can be very large (small) depending upon the parameter space 
under consideration and parameters such as $f_a$ and $|e_Q|$. 
We can measure the lifetimes of these charged massive particles at collider according to their decay lengths ($c\tau$).
These lifetime measurements at colliders can be divided into three scenarios depending upon 
the decay length ($c\tau_{\tilde \tau_1}$) and the size
of the detector $L$ which is typically $L< 0.1$ km. In the first scenario $c\tau_{\tilde\tau_1} \ll O(cm)$, with very 
short lifetime $\tau_{\ttau_1} \ll 10^{-10}s$. In the second
scenario the lifetime is supposed to be $O(cm) \lesssim c\tau_{\tilde \tau_1}\lesssim L$,
while in the third scenario we have $c\tau_{\tilde \tau_1}\gtrsim \,L$. In Fig.~\ref{fa8}, we present
 the lifetime (in the left panel) and 
the decay length (in the right panel) of staus for our data. Here we have used the minimal value of $f_a$ 
with two different values of $|e_Q|$ to see 
what are the minimal lifetimes and decay lengths in our present data. In this
 figure, the aqua points correspond to $f_a=7\times 10^{8}$ and $|e_Q|=1/3$, and the purple points 
represent $f_a=7\times 10^{8}$ and $|e_Q|=1$.  In the left panel, we see that 
for aqua and purple points the lifetimes are respectively $\tau_{\tilde\tau_1}$ between $10^{-5}$~s to $10^{-4}$~s 
and $10^{-7}$~s to $10^{-6}$~s. In right panel, the minimal decay length for purple points ($|e_Q|=1$) is less than 0.1 km, 
while for aqua points ($|e_Q|=1/3$) it is about 5 km.  This figure shows that with heavy stau masses, 
the decay length of stau can be within the detector $L< 0.1$ km. If such staus decay to axinos, it is possible that we may observe 
LSP axino scenario with $|e_Q|=1$.  Moreover, it is estimated 
in Refs.~\cite{Ishiwata:2008tp,Kaneko:2008re} that for a relatively small lifetime $10^{-3}$ to $10^{-5}$ s, 
we can still observe a 
substantial number of in-flight decays of staus within the detector.
It is expected that the staus with large lifetimes may loose their energies through ionization and slowed down. 
These staus can be stopped within the main
detector~\cite{Martyn:2006as,Asai:2009ka,Pinfold:2010aq}, or an additional dedicated stopping-detector can be used~\cite{Goity:1993ih,Hamaguchi:2004df,Hamaguchi:2006vu}. In this case, it is possible to study the subsequent decays 
of trapped staus by recording their stopping points/times via those devices.
Furthermore, there is a proposal to place a water tank around the detector~\cite{Feng:2004yi} 
or use the surrounding rock as a detector for the study of trapped stau decays~\cite{DeRoeck:2005bw}.  
In this paper we do not study the 2-body and 3-body decays of stau NLSP for our data at colliders and 
leave it as a future project. Such studies can be found in Ref.~\cite{Freitas:2011fx}. 
The collider phenomenology with the directly produced long-lived staus within the MSSM and 
calculations for dominant cross section contributions to stau pair productions are reported 
in Ref.~\cite{Lindert:2011td}. In fact, the collider searches for charged long-lived particles have
 been performed at the LEP~\cite{Barate:1997dr,Abreu:2000tn,Achard:2001qw,Abbiendi:2003yd} , HERA~~\cite{Aktas:2004pq}, the Tevatron~~\cite{Abazov:2007ht,Aaltonen:2009kea,Abazov:2011pf}, and the LHC~~\cite{Aad:2012pra,Aad:2013gva}. Recently,
 the ATLAS Collaboration \cite{ATLAS:2014fka} has reported that the 
long-lived stau, in the SSMs with gauge-mediated supersymmetry breaking 
(GMSB), is excluded up to its masses between 440 and 385 GeV for $\tan\beta$ between 10 and 50, with a 290 GeV limit in the case where only the direct stau 
production is considered. In the context of the simplified Lepto-SUSY models, where sleptons are stable and have a mass of 300 GeV, squark and gluino masses 
are excluded up to a mass of 1500 and 1360 GeV, respectively. In addition, 
the stau mass below 500 (339) GeV is excluded for the direct+indirect 
(direct only) production by the CMS Collaboration~\cite{Chatrchyan:2013oca}. While the LHCb 
Collaboration has looked for stau with mass 124 and 309 GeV without any success 
and has put limits on di-stau production cross section~\cite{LHCb:21 07 2014asa}. 
Because the GMSB scenario is assumed in the latest studies of CHAPMs at the LHC, 
these stau mass bounds are for a different kind of the SSMs. The implications 
of these results on the No-Scale MSSM 
would necessitate a detailed study. At the moment, just taking these stau mass limits, 
it seems to us that our stau NLSP  solutions are heavy enough to evade the LHC bounds.  

 We display three benchmark points in Table~\ref{table1} with gaugino masses about 1866~GeV,
2725~GeV, and 4589~GeV as well as Higgs boson masses about 123~GeV, 125~GeV,
and 127~GeV, respectively.
 Point 1 represents the scenario where we have data satisfying all the current constraints
 discussed previously, as well as have $M_{1/2}\lesssim$ 2000 GeV and Higgs mass around 123 GeV. 
Point 2 depicts the scenario where we have $M_{1/2}$ between 2000 GeV to 3000 GeV and 
Higgs mass mass around 125 GeV. Point 3 is an example of points with $M_{1/2}\gtrsim $ 3000 GeV
 and Higgs mass around 127 GeV.  We can see that in all these Points the sparticle spectra are heavy. 
For instance, for Point 1, we see that the masses for the first two families of squarks  are 
slightly above 3000 GeV, while the stops are around 3000 GeV and gluino mass 
is around 3895 GeV. Similarly, the masses for the left-handed slepton of the
first two families are heavier than 1000 GeV but the right-handed slepton masses are about 679 GeV, 
while $m_{\tilde \tau_{1,2}}$ are 587 GeV and 1184 GeV, respectively. Electroweakinos are also 
heavy and in the range of [800,2000] GeV. For Points 2 and 3, the
particle  spectra are even more heavy. The masses for the first two families of squarks are above 4700 GeV 
 and  7600 GeV respectively for Points 2 and 3. Similarly the light stops are about 4000 GeV 
and 6400 GeV. The left-handed slepton masses for the first two families are respectively 1739 GeV and 2890 GeV 
for Points 2 and 3  while the right-handed slepton masses are 987 GeV and 1653 GeV. The NLSP staus are 841 GeV and 1374 GeV for Points 2 and 3, respectively. Such heavy spectra can not be probed at the 14 TeV LHC,
 which will provide a strong motivation for 33 TeV and 100 TeV proton-proton colliders. 
In Ref.~\cite{cern_note1}, it was shown that the squarks/gluinos of 2.5 TeV, 3 TeV and 6 TeV may be probed by
the LHC14, High Luminosity (HL)LHC14 and High Energy (HE) LHC33, respectively. This
clearly shows that our models have testable predictions. Moreover, in the future if we have collider
facility with even higher energy, we will be able to probe over even larger values of sparticle
masses.



\section{Discussions and Conclusion}
\label{summary}

 We emphasized that the electroweak fine-tuning problem in the SSMs is mainly because of 
 the high energy definition of fine-tuning measure. We proposed the super-natural supersymmetry 
with the order one high energy fine-tuning measure naturally since
 all the mass parameters in the SSMs arise from a single supersymmetry breaking parameter. 
With the numerical calculations of the fine-tuning measures, 
we showed explicitly that we indeed do not have the supersymmetry electroweak fine-tuning problem 
in the MSSM with no-scale supergravity and Giudice-Masiero (GM) mechanism.
We studied various aspects in the No-Scale MSSM, and found that the $Z$-boson mass, 
the $\mu(GUT)$ parameter and the sparticle
spectra can be given as functions of the universal gaugino mass parameter $M_{1/2}$. 
Introducing a non-thermally generated axino LSP, we estimated the lifetime of stau NSLP by calculating its 
2-body body and 3-body decays to the axino LSP for various values of axion decay constant $f_a$.
In particular, our stau NSLP solutions satisfy all the current experimental 
bounds, such as sparticle mass bounds, B-physics bounds, 
Higgs mass bounds, cosmological bounds, and
the bounds on long-lived charge particles at the LHC.
\begin{table}[b]\hspace{-1.0cm}
\centering
\begin{tabular}{|c|ccc|}
\hline
\hline
                 & Point 1 & Point 2 & Point 3 \\

\hline
$M_{1/2}$  & 1865.849  &  2725.144   & 4588.77  \\
 $m_{0}$    & 0.0       &  0.0        & 0.0    \\
$A_{0}$        & 0.0       &  0.0        & 0.0   \\
$\tan\beta$      & 28.129    &  30.594     & 33.822  \\
\hline
$B_{0}\tiny\text{(GUT)}$  & 0.599    &  0.251    & -0.716  \\
$\mu \tiny\text{(GUT)}$    & 1991     &  2788     & 4431 \\
$B_{0}\tiny\text{(EWSB)}$  & 92       &  118      & 4292 \\
$\mu \tiny\text{(EWSB)}$   & 1970     &  2735     & 166 \\
\hline

\hline
$m_h$            & 123  & 125 & 127 \\
$m_H$            & 2054 & 2816 & 4321  \\
$m_A$            & 2054 & 2815 & 4321 \\
$m_{H^{\pm}}$    & 2056 & 2816 & 4322 \\

\hline
$m_{\tilde{\chi}^0_{1,2}}$
                 & 820, 1520    & 1219, 2238    & 2100, 3802    \\

$m_{\tilde{\chi}^0_{3,4}}$
                 & 1978, 1985    & 2747, 2753    & 4310, 4317     \\

$m_{\tilde{\chi}^{\pm}_{1,2}}$
                 & 1520, 1985    & 2238, 2753    & 3802, 4317     \\
\hline
$m_{\tilde{g}}$  & 3895 & 5546  & 9030 \\
\hline $m_{ \tilde{u}_{L,R}}$
                 & 3505, 3358    & 4968, 4746   & 8035, 7604       \\
$m_{\tilde{t}_{1,2}}$
                 & 2786, 3225   & 3957, 4554     & 6411, 7343       \\
\hline
$m_{ \tilde{d}_{L,R}}$
                 & 3506, 3356    & 4968, 4717     & 8035, 7604       \\
$m_{\tilde{b}_{1,2}}$
                & 3205, 3247    & 4530, 4574     & 7268, 7364       \\
$m_{\tilde{\nu}_{1,2}}$
                 &1199  & 1737  &  2889\\
$m_{\tilde{\nu}_{3}}$
                 & 1177  & 1701  &  2819\\
\hline
$m_{ \tilde{e}_{L,R}}$
                & 1202, 679   & 1739, 987    & 2890, 1653        \\
$m_{\tilde{\tau}_{1,2}}$
                & 587,1184    & 841, 1705    & 1374, 2822        \\


\hline
\hline
\end{tabular}
\caption{All the masses are in this table are in units of GeV and $\mu>0$.
All this points satisfy  the sparticle mass, B-physics constraints described in Section~\ref{sec:scan}.
Points 1, 2 and 3 represent parameter space with $m_{h}$ = 123 GeV, 125 GeV and 127 GeV respectively.
}
\label{table1}
\end{table}

\section*{Acknowledgments}
We would like to thank Laura Covi, Ayres Freitas and Frank D. Steffen for very useful discussions. 
This research was supported in part by the 
Natural Science Foundation of China under grant numbers 10821504, 11075194, 11135003, 11275246, and 11475238,  
and by the National
Basic Research Program of China (973 Program) under grant number 2010CB833000 (TL), and by
 the DOE grant DE-FG03-95-ER-40917 (DVN).



\begin{thebibliography}{100}
\bibitem{ATLAS}
  G.~Aad {\it et al.}  [ATLAS Collaboration],
  Phys.\ Lett.\ B {\bf 716}, 1 (2012)
  [arXiv:1207.7214 [hep-ex]].


\bibitem{CMS}
  S.~Chatrchyan {\it et al.}  [CMS Collaboration],
  Phys.\ Lett.\ B {\bf 716}, 30 (2012)
  [arXiv:1207.7235 [hep-ex]].
\bibitem{Carena:2011aa}
  M.~Carena, S.~Gori, N.~R.~Shah and C.~E.~M.~Wagner,
  JHEP {\bf 1203}, 014 (2012)
  [arXiv:1112.3336 [hep-ph]].
\bibitem{Ellwanger:2011aa}
  U.~Ellwanger,
  JHEP {\bf 1203}, 044 (2012)
  [arXiv:1112.3548 [hep-ph]].


\bibitem{Kang:2012sy}
  Z.~Kang, J.~Li and T.~Li,
  JHEP {\bf 1211}, 024 (2012)
  [arXiv:1201.5305 [hep-ph]].

\bibitem{Cao:2012fz}
  J.~J.~Cao, Z.~X.~Heng, J.~M.~Yang, Y.~M.~Zhang and J.~Y.~Zhu,
  JHEP {\bf 1203}, 086 (2012)
  [arXiv:1202.5821 [hep-ph]].
\bibitem{Aad:2014wea}
  G.~Aad {\it et al.}  [ATLAS Collaboration],
  arXiv:1405.7875 [hep-ex].
\bibitem{Chatrchyan:2013wxa}
  S.~Chatrchyan {\it et al.}  [CMS Collaboration],
  Phys.\ Lett.\ B {\bf 725}, 243 (2013)
  [arXiv:1305.2390 [hep-ex]].
\bibitem{Kitano:2005wc}
  R.~Kitano and Y.~Nomura,
  Phys.\ Lett.\ B {\bf 631}, 58 (2005)
  [hep-ph/0509039];
  Phys.\ Rev.\ D {\bf 73}, 095004 (2006)
  [hep-ph/0602096].

\bibitem{Papucci:2011wy}
  M.~Papucci, J.~T.~Ruderman and A.~Weiler,
  arXiv:1110.6926 [hep-ph].

\bibitem{Baer:2012mv}
  H.~Baer, V.~Barger, P.~Huang, D.~Mickelson, A.~Mustafayev and X.~Tata,
  Phys.\ Rev.\ D {\bf 87}, no. 3, 035017 (2013)
  [arXiv:1210.3019 [hep-ph]].
\bibitem{Ellis:1986yg}
  J.~R.~Ellis, K.~Enqvist, D.~V.~Nanopoulos and F.~Zwirner,
  Mod.\ Phys.\ Lett.\ A {\bf 1}, 57 (1986).
\bibitem{Barbieri:1987fn}
  R.~Barbieri and G.~F.~Giudice,
  Nucl.\ Phys.\ B {\bf 306}, 63 (1988).
\bibitem{Leggett:2014mza}
  T.~Leggett, T.~Li, J.~A.~Maxin, D.~V.~Nanopoulos and J.~W.~Walker,
  arXiv:1403.3099 [hep-ph].
\bibitem{Leggett:2014hha}
  T.~Leggett, T.~Li, J.~A.~Maxin, D.~V.~Nanopoulos and J.~W.~Walker,
  Phys.\ Lett.\ B {\bf 740}, 66 (2015)
  [arXiv:1408.4459 [hep-ph]].



\bibitem{F-SU5} 
S. M. Barr,
Phys.\ Lett.\ B {\bf 112}, 219 (1982);
J.~P.~Derendinger, J.~E.~Kim and D.~V.~Nanopoulos,
Phys.\ Lett.\ B {\bf 139}, 170 (1984);
  I.~Antoniadis, J.~R.~Ellis, J.~S.~Hagelin and D.~V.~Nanopoulos,
  Phys.\ Lett.\  B {\bf 194}, 231 (1987).

\bibitem{Jiang:2006hf}
  J.~Jiang, T.~Li and D.~V.~Nanopoulos,
  Nucl.\ Phys.\  B {\bf 772}, 49 (2007).
  
\bibitem{Jiang:2009zza}
  J.~Jiang, T.~Li, D.~V.~Nanopoulos and D.~Xie,
  Phys.\ Lett.\  B {\bf 677}, 322 (2009);
  Nucl.\ Phys.\  B {\bf 830}, 195 (2010).

\bibitem{Li:2010ws} 
  T.~Li, J.~A.~Maxin, D.~V.~Nanopoulos and J.~W.~Walker,
  Phys.\ Rev.\ D {\bf 83}, 056015 (2011)
  [arXiv:1007.5100 [hep-ph]];
  Phys.\ Lett.\ B {\bf 699}, 164 (2011)
  [arXiv:1009.2981 [hep-ph]].


\bibitem{Cremmer:1983bf}
  E.~Cremmer, S.~Ferrara, C.~Kounnas and D.~V.~Nanopoulos,
  Phys.\ Lett.\  B {\bf 133}, 61 (1983);
J.~R.~Ellis, A.~B.~Lahanas, D.~V.~Nanopoulos and K.~Tamvakis,
  Phys.\ Lett.\  B {\bf 134}, 429 (1984);
J.~R.~Ellis, C.~Kounnas and D.~V.~Nanopoulos,
  Nucl.\ Phys.\  B {\bf 241}, 406 (1984);
  Nucl.\ Phys.\  B {\bf 247}, 373 (1984);
A.~B.~Lahanas and D.~V.~Nanopoulos,
  Phys.\ Rept.\  {\bf 145}, 1 (1987).



\bibitem{Giudice:1988yz}
  G.~F.~Giudice and A.~Masiero,
  Phys.\ Lett.\ B {\bf 206}, 480 (1988).
\bibitem{Raffelt:2006cw}
G.~G. Raffelt,
\newblock Lect. Notes Phys. {\bf 741}, 51 (2008), hep-ph/0611350.

\bibitem{Ferrara:1994kg}
  S.~Ferrara, C.~Kounnas and F.~Zwirner,
  Nucl.\ Phys.\ B {\bf 429}, 589 (1994)
  [Erratum-ibid.\ B {\bf 433}, 255 (1995)]
  [hep-th/9405188].
\bibitem{Aad:2014nua}
  G.~Aad {\it et al.}  [ATLAS Collaboration],
  JHEP {\bf 1404}, 169 (2014)
  [arXiv:1402.7029 [hep-ex]].

\bibitem{Aad:2014vma}
  G.~Aad {\it et al.}  [ATLAS Collaboration],
  JHEP {\bf 1405}, 071 (2014)
  [arXiv:1403.5294 [hep-ex]].


\bibitem{CMS-PAS-SUS-13-006}
{\bf CMS} Collaboration,
``Search for electroweak production of charginos, neutralinos, and sleptons using leptonic final states in pp
collisions at 8 TeV.''
[CMS-PAS-SUS-13-006]



\bibitem{Li:2014dna}
  T.~Li and S.~Raza,
  arXiv:1409.3930 [hep-ph].





\bibitem{Djouadi:2002ze} 
  A.~Djouadi, J.~L.~Kneur and G.~Moultaka,
  Comput.\ Phys.\ Commun.\  {\bf 176}, 426 (2007)
  [hep-ph/0211331].

\bibitem{ATLAS:2014wva} 
  [ATLAS and CDF and CMS and D0 Collaborations],
  arXiv:1403.4427 [hep-ex].


\bibitem{PDG} 
  K.~.A.~Olive {\it et al.}  [Particle Data Group],
 Chin. Phys. C, 38, 090001 (2014).



\bibitem{Ade:2013zuv} 
  P.~A.~R.~Ade {\it et al.}  [Planck Collaboration],
  Astron.\ Astrophys.\  {\bf 571}, A16 (2014)
  [arXiv:1303.5076 [astro-ph.CO]].
  P.~A.~R.~Ade {\it et al.}  [Planck Collaboration],
  Astron.\ Astrophys.\  {\bf 571}, A15 (2014)
  [arXiv:1303.5075 [astro-ph.CO]].
  P.~A.~R.~Ade {\it et al.}  [Planck Collaboration],
  Astron.\ Astrophys.\  {\bf 571}, A1 (2014)
  [arXiv:1303.5062 [astro-ph.CO]].

\bibitem{Hinshaw:2012aka} 
  G.~Hinshaw {\it et al.}  [WMAP Collaboration],
  Astrophys.\ J.\ Suppl.\  {\bf 208}, 19 (2013)
  [arXiv:1212.5226 [astro-ph.CO]].

\bibitem{Goldberg:1983nd} 
  H.~Goldberg,
  Phys.\ Rev.\ Lett.\  {\bf 50}, 1419 (1983)
  [Erratum-ibid.\  {\bf 103}, 099905 (2009)].

\bibitem{Ellis:1983ew} 
  J.~R.~Ellis, J.~S.~Hagelin, D.~V.~Nanopoulos, K.~A.~Olive and M.~Srednicki,
  Nucl.\ Phys.\ B {\bf 238}, 453 (1984).


\bibitem{Jungman:1995df} 
  G.~Jungman, M.~Kamionkowski and K.~Griest,
  Phys.\ Rept.\  {\bf 267}, 195 (1996)
  [hep-ph/9506380].

\bibitem{Bertone:2004pz} 
  G.~Bertone, D.~Hooper and J.~Silk,
  Phys.\ Rept.\  {\bf 405}, 279 (2005)
  [hep-ph/0404175].
\bibitem{Schmaltz:2000gy}
  M.~Schmaltz and W.~Skiba,
  Phys.\ Rev.\ D {\bf 62}, 095005 (2000)
  [hep-ph/0001172].

\bibitem{Ellis:2010jb}
  J.~Ellis, A.~Mustafayev and K.~A.~Olive,
  Eur.\ Phys.\ J.\ C {\bf 69}, 219 (2010)
  [arXiv:1004.5399 [hep-ph]].







\bibitem{Kim:1984yn} 
  J.~E.~Kim, A.~Masiero and D.~V.~Nanopoulos,
  Phys.\ Lett.\ B {\bf 139}, 346 (1984).

\bibitem{Bonometto:1993fx}
S.A. Bonometto, F.~Gabbiani, A.~Masiero, Phys. Rev. \textbf{D49}, 3918 (1994),
  \texttt{hep-ph/9305237}

\bibitem{Covi:1999ty}
L.~Covi, J.E. Kim, L.~Roszkowski, Phys. Rev. Lett. \textbf{82}, 4180 (1999),
  \texttt{hep-ph/9905212}

\bibitem{Covi:2001nw}
L.~Covi, H.B. Kim, J.E. Kim, L.~Roszkowski, JHEP \textbf{05}, 033 (2001),
  \texttt{hep-ph/0101009}

\bibitem{Covi:2004rb}
L.~Covi, L.~Roszkowski, R.~Ruiz~de Austri, M.~Small, JHEP \textbf{06}, 003
  (2004), \texttt{hep-ph/0402240}

\bibitem{Brandenburg:2004du}
A.~Brandenburg, F.D. Steffen, JCAP \textbf{0408}, 008 (2004),
  \texttt{hep-ph/0405158}

\bibitem{Steffen:2008qp}
F.D. Steffen, Eur. Phys. J. \textbf{C59}, 557 (2009), \texttt{0811.3347}

\bibitem{Baer:2008yd}
H.~Baer, M.~Haider, S.~Kraml, S.~Sekmen, H.~Summy, JCAP \textbf{0902}, 002
  (2009), \texttt{0812.2693}

\bibitem{Peccei:1977hh} 
  R.~D.~Peccei and H.~R.~Quinn,
  Phys.\ Rev.\ Lett.\  {\bf 38}, 1440 (1977).

\bibitem{Peccei:1977ur} 
  R.~D.~Peccei and H.~R.~Quinn,
  Phys.\ Rev.\ D {\bf 16}, 1791 (1977).

\bibitem{Tamvakis:1982mw} 
  K.~Tamvakis and D.~Wyler,
  Phys.\ Lett.\ B {\bf 112}, 451 (1982).
\bibitem{Nieves:1985fq} 
  J.~F.~Nieves,
  Phys.\ Rev.\ D {\bf 33}, 1762 (1986).
\bibitem{Rajagopal:1990yx} 
  K.~Rajagopal, M.~S.~Turner and F.~Wilczek,
  Nucl.\ Phys.\ B {\bf 358}, 447 (1991).
\bibitem{Goto:1991gq} 
  T.~Goto and M.~Yamaguchi,
  Phys.\ Lett.\ B {\bf 276}, 103 (1992).
  E.~J.~Chun, J.~E.~Kim and H.~P.~Nilles,
  Phys.\ Lett.\ B {\bf 287}, 123 (1992)
  [hep-ph/9205229].
  E.~J.~Chun and A.~Lukas,
  Phys.\ Lett.\ B {\bf 357} (1995) 43
  [hep-ph/9503233].
\bibitem{Drees:1990yw}
M.~Drees and X.~Tata,
\newblock Phys. Lett. {\bf B252}, 695 (1990).

\bibitem{Nisati:1997gb}
A.~Nisati, S.~Petrarca, and G.~Salvini,
\newblock Mod. Phys. Lett. {\bf A12}, 2213 (1997), hep-ph/9707376.

\bibitem{Ambrosanio:1997rv}
S.~Ambrosanio, G.~D. Kribs, and S.~P. Martin,
\newblock Phys. Rev. {\bf D56}, 1761 (1997), hep-ph/9703211.

\bibitem{Feng:1997zr}
J.~L. Feng and T.~Moroi,
\newblock Phys. Rev. {\bf D58}, 035001 (1998), hep-ph/9712499.

\bibitem{Fairbairn:2006gg}
M.~Fairbairn {\em et~al.},
\newblock Phys. Rept. {\bf 438}, 1 (2007), hep-ph/0611040.

\bibitem{Martin:1998vb}
S.~P. Martin and J.~D. Wells,
\newblock Phys. Rev. {\bf D59}, 035008 (1999), hep-ph/9805289.
\bibitem{LHCb:21 07 2014asa} 
  The LHCb Collaboration [LHCb Collaboration],
  LHCb-CONF-2014-001, CERN-LHCb-CONF-2014-001.

\bibitem{Endo:2011uw} 
  M.~Endo, K.~Hamaguchi and K.~Nakaji,
  arXiv:1105.3823 [hep-ph].


\bibitem{Ito:2010un} 
  T.~Ito,
  Phys.\ Lett.\ B {\bf 699}, 151 (2011)
  [arXiv:1012.1318 [hep-ph]].


\bibitem{Ito:2010xj} 
  T.~Ito and T.~Moroi,
  Phys.\ Lett.\ B {\bf 694}, 349 (2011)
  [arXiv:1007.3060 [hep-ph]].


\bibitem{Kitano:2010tt} 
  R.~Kitano and M.~Nakamura,
  Phys.\ Rev.\ D {\bf 82}, 035007 (2010)
  [arXiv:1006.2904 [hep-ph]].


\bibitem{Ito:2009xy} 
  T.~Ito, R.~Kitano and T.~Moroi,
  JHEP {\bf 1004}, 017 (2010)
  [arXiv:0910.5853 [hep-ph]].


\bibitem{Pradler:2009mt} 
  J.~Pradler,
  arXiv:0909.3429 [hep-ph].

\bibitem{Jittoh:2009zzc} 
  T.~Jittoh, K.~Kohri, M.~Koike, J.~Sato, T.~Shimomura and M.~Yamanaka,
  AIP Conf.\ Proc.\  {\bf 1115}, 285 (2009).


\bibitem{Yamanaka:2009zzb} 
  M.~Yamanaka, T.~Jittoh, K.~Kohri, M.~Koike, J.~Sato and T.~Shimomura,
  PoS EPS {\bf -HEP2009}, 115 (2009).


\bibitem{Kohri:2008zza} 
  K.~Kohri, T.~Jittoh, M.~Koike, J.~Sato, T.~Shimomura and M.~Yamanaka,
  AIP Conf.\ Proc.\  {\bf 1006}, 126 (2008).


\bibitem{Jittoh:2007fr} 
  T.~Jittoh, K.~Kohri, M.~Koike, J.~Sato, T.~Shimomura and M.~Yamanaka,
  Phys.\ Rev.\ D {\bf 76}, 125023 (2007)
  [arXiv:0704.2914 [hep-ph]].


\bibitem{Ibarra:2006sz} 
  A.~Ibarra and S.~Roy,
  JHEP {\bf 0705}, 059 (2007)
  [hep-ph/0606116].
\bibitem{Tarem:2005gxa} 
  S.~Tarem, S.~Bressler, E.~Duchovni and L.~Levinson,
  ATL-PHYS-PUB-2005-022, CERN-ATL-PHYS-PUB-2005-022, ATL-COM-PHYS-2005-051.
\bibitem{Freitas:2009jb} 
  A.~Freitas, F.~D.~Steffen, N.~Tajuddin and D.~Wyler,
  Phys.\ Lett.\ B {\bf 682}, 193 (2009)
  [arXiv:0909.3293 [hep-ph]];K.~Jedamzik, Phys. Rev. {\bf D74}, 103509 (2006), hep-ph/0604251; R.~H. Cyburt {\em et~al.},
 JCAP {\bf 0910}, 021 (2009), 0907.5003; M.~H. Reno and D.~Seckel,
 Phys. Rev. {\bf D37}, 3441 (1988); S.~Dimopoulos, R.~Esmailzadeh, L.~J. Hall, and G.~D. Starkman,
 Astrophys. J. {\bf 330}, 545 (1988); S.~Dimopoulos, R.~Esmailzadeh, L.~J. Hall, and G.~D. Starkman,
 Nucl. Phys. {\bf B311}, 699 (1989); K.~Kohri, Phys. Rev. {\bf D64}, 043515 (2001), astro-ph/0103411; M.~Kawasaki, K.~Kohri, and T.~Moroi,
Phys. Rev. {\bf D71}, 083502 (2005), astro-ph/0408426; V.~Simha and G.~Steigman, JCAP {\bf 0806}, 016 (2008), 0803.3465;
Y.~I. Izotov and T.~X. Thuan, Astrophys. J. {\bf 710}, L67 (2010), 1001.4440; 
E.~Aver, K.~A. Olive, and E.~D. Skillman, JCAP {\bf 1005}, 003 (2010), 1001.5218; E.~Holtmann, M.~Kawasaki, K.~Kohri, and T.~Moroi,
Phys. Rev. {\bf D60}, 023506 (1999), hep-ph/9805405; R.~H. Cyburt, J.~R. Ellis, B.~D. Fields, and K.~A. Olive, Phys. Rev. {\bf D67}, 103521 (2003), astro-ph/0211258; M.~Kawasaki, K.~Kohri, and T.~Moroi, Phys. Lett. {\bf B625}, 7 (2005), astro-ph/0402490; 
M.~Kawasaki, K.~Kohri, T.~Moroi, and A.~Yotsuyanagi, Phys. Rev. {\bf D78}, 065011 (2008), 0804.3745.

\bibitem{Freitas:2009fb}
A.~Freitas, F.~D. Steffen, N.~Tajuddin, and D.~Wyler,
Phys. Lett. {\bf B679}, 270 (2009), 0904.3218.

\bibitem{pospelov}
M.~Pospelov, Phys. Rev. Lett. {\bf 98}, 231301 (2007), hep-ph/0605215;
M.~Pospelov, {\em Bridging the primordial A=8 divide with Catalyzed Big Bang Nucleosynthesis}
(2007), 0712.0647; M.~Pospelov, J.~Pradler, and F.~D. Steffen, JCAP {\bf 0811}, 020 (2008), 0807.4287.


\bibitem{cyburt}
R.~H. Cyburt, J.~R. Ellis, B.~D. Fields, and K.~A. Olive, Phys. Rev. {\bf D67}, 103521 (2003), astro-ph/0211258;
M.~Asplund, D.~L. Lambert, P.~E. Nissen, F.~Primas, and V.~V. Smith, Astrophys. J. {\bf 644}, 229 (2006), astro-ph/0510636;
K.~Jedamzik, JCAP {\bf 0803}, 008 (2008), 0710.5153v2.

\bibitem{Polisensky:2010rw}
E.~Polisensky and M.~Ricotti,
\newblock Phys. Rev. {\bf D83}, 043506 (2011), 1004.1459; A.~V. Maccio' and F.~Fontanot,
{\em How cold is Dark Matter? Constraints from Milky Way Satellites} (2009), 0910.2460;
J.~J. Dalcanton and C.~J. Hogan, Astrophys. J. {\bf 561}, 35 (2001), astro-ph/0004381;
D.~Boyanovsky, H.~J. de~Vega, and N.~Sanchez, Phys. Rev. {\bf D77}, 043518 (2008), 0710.5180;
V.~K. Narayanan, D.~N. Spergel, R.~Dave, and C.-P. Ma, (2000), astro-ph/0005095;
M.~Viel, J.~Lesgourgues, M.~G. Haehnelt, S.~Matarrese, and A.~Riotto, Phys. Rev. {\bf D71}, 063534 (2005), astro-ph/0501562;
U.~Seljak, A.~Makarov, P.~McDonald, and H.~Trac, Phys. Rev. Lett. {\bf 97}, 191303 (2006), astro-ph/0602430;
M.~Viel, J.~Lesgourgues, M.~G. Haehnelt, S.~Matarrese, and A.~Riotto, Phys. Rev. Lett. {\bf 97}, 071301 (2006), astro-ph/0605706;
M.~Viel {\em et~al.}, Phys. Rev. Lett. {\bf 100}, 041304 (2008), 0709.0131.


\bibitem{Benhenni:2011yt} 
  A.~Benhenni, J.~L.~Kneur, G.~Moultaka and S.~Bailly,
  Phys.\ Rev.\ D {\bf 84}, 075015 (2011)
  [arXiv:1106.6325 [hep-ph]].
\bibitem{Heisig:2013sva} 
  J.~Heisig,
  JCAP {\bf 1404}, 023 (2014)
  [arXiv:1310.6352 [hep-ph]].
\bibitem{Panotopoulos:2008cs} 
  G.~Panotopoulos,
  Phys.\ Lett.\ B {\bf 671}, 327 (2009)
  [arXiv:0812.3987 [hep-ph]].
\bibitem{Feng:2004mt} 
  J.~L.~Feng, S.~Su and F.~Takayama,
  Phys.\ Rev.\ D {\bf 70}, 075019 (2004)
  [hep-ph/0404231].
\bibitem{micromegas}
G.~B\'elanger, F.~Boudjema, and A.~Pukhov, {\em micrOMEGAs: A Package for
  Calculation of Dark Matter Properties in Generic Model of Particle
  Interaction}, ch.~12, pp.~739--790.
\newblock World Scientific, 2012.
\newblock
  \href{http://www.arXiv.org/abs/http://www.worldscientific.com/doi/pdf/10.1142/9789814390163\_0012}{{\tt
  http://www.worldscientific.com/doi/pdf/10.1142/9789814390163\_0012}}.


%
\bibitem{mamoudi} D.~Eriksson, F.~Mahmoudi and O.~Stal,
\jhep{0811}{2008}{035}.



\bibitem{Aaij:2012nna}
  R.~Aaij {\it et al.}  [LHCb Collaboration],
  Phys.\ Rev.\ Lett.\  {\bf 110}, 021801 (2013).
\bibitem{Amhis:2012bh}
  Y.~Amhis {\it et al.}  [Heavy Flavor Averaging Group Collaboration],
  arXiv:1207.1158 [hep-ex].


\bibitem{Asner:2010qj}
  D.~Asner {\it et al.}  [Heavy Flavor Averaging Group Collaboration],
  arXiv:1010.1589 [hep-ex].







\bibitem{Brandenburg:2005he} 
  A.~Brandenburg, L.~Covi, K.~Hamaguchi, L.~Roszkowski and F.~D.~Steffen,
  Phys.\ Lett.\ B {\bf 617}, 99 (2005)
  [hep-ph/0501287].
\bibitem{Freitas:2011fx} 
  A.~Freitas, F.~D.~Steffen, N.~Tajuddin and D.~Wyler,
  JHEP {\bf 1106}, 036 (2011)
  [arXiv:1105.1113 [hep-ph]].
\bibitem{Kim:1979if} 
  J.~E.~Kim,
  Phys.\ Rev.\ Lett.\  {\bf 43}, 103 (1979).
\bibitem{Shifman:1979if} 
  M.~A.~Shifman, A.~I.~Vainshtein and V.~I.~Zakharov,
  Nucl.\ Phys.\ B {\bf 166}, 493 (1980).
\bibitem{'tHooft:1978xw} 
  G.~'t Hooft and M.~J.~G.~Veltman,
  Nucl.\ Phys.\ B {\bf 153}, 365 (1979).

\bibitem{Hahn:1998yk} 
  T.~Hahn and M.~Perez-Victoria,
  Comput.\ Phys.\ Commun.\  {\bf 118}, 153 (1999)
  [hep-ph/9807565].
\bibitem{Ishiwata:2008tp}
K.~Ishiwata, T.~Ito, and T.~Moroi,
\newblock Phys. Lett. {\bf B669}, 28 (2008), 0807.0975.

\bibitem{Kaneko:2008re}
S.~Kaneko, J.~Sato, T.~Shimomura, O.~Vives, and M.~Yamanaka,
\newblock Phys. Rev. {\bf D78}, 116013 (2008), 0811.0703.
\bibitem{Martyn:2006as}
H.~U. Martyn,
\newblock Eur. Phys. J. {\bf C48}, 15 (2006), hep-ph/0605257.

\bibitem{Asai:2009ka}
S.~Asai, K.~Hamaguchi, and S.~Shirai,
\newblock Phys. Rev. Lett. {\bf 103}, 141803 (2009), 0902.3754.

\bibitem{Pinfold:2010aq}
J.~Pinfold and L.~Sibley,
\newblock Phys. Rev. {\bf D83}, 035021 (2011), 1006.3293.

\bibitem{Goity:1993ih}
J.~L. Goity, W.~J. Kossler, and M.~Sher,
\newblock Phys. Rev. {\bf D48}, 5437 (1993), hep-ph/9305244.

\bibitem{Hamaguchi:2004df}
K.~Hamaguchi, Y.~Kuno, T.~Nakaya, and M.~M. Nojiri,
\newblock Phys. Rev. {\bf D70}, 115007 (2004), hep-ph/0409248.
\bibitem{Hamaguchi:2006vu}
K.~Hamaguchi, M.~M. Nojiri, and A.~de~Roeck,
\newblock JHEP {\bf 03}, 046 (2007), hep-ph/0612060.
\bibitem{Feng:2004yi}
J.~L. Feng and B.~T. Smith,
\newblock Phys. Rev. {\bf D71}, 015004 (2005), hep-ph/0409278.
\bibitem{DeRoeck:2005bw}
A.~De~Roeck {\em et~al.},
\newblock Eur. Phys. J. {\bf C49}, 1041 (2007), hep-ph/0508198.

\bibitem{Lindert:2011td} 
  J.~M.~Lindert, F.~D.~Steffen and M.~K.~Trenkel,
  JHEP {\bf 1108}, 151 (2011)
  [arXiv:1106.4005 [hep-ph]].
\bibitem{Barate:1997dr} 
  R.~Barate {\it et al.}  [ALEPH Collaboration],
  Phys.\ Lett.\ B {\bf 405}, 379 (1997)
  [hep-ex/9706013].

\bibitem{Abreu:2000tn} 
  P.~Abreu {\it et al.}  [DELPHI Collaboration],
  Phys.\ Lett.\ B {\bf 478}, 65 (2000)
  [hep-ex/0103038].


\bibitem{Achard:2001qw} 
  P.~Achard {\it et al.}  [L3 Collaboration],
  Phys.\ Lett.\ B {\bf 517}, 75 (2001)
  [hep-ex/0107015].

\bibitem{Abbiendi:2003yd} 
  G.~Abbiendi {\it et al.}  [OPAL Collaboration],
  Phys.\ Lett.\ B {\bf 572}, 8 (2003)
  [hep-ex/0305031].

\bibitem{Aktas:2004pq} 
  A.~Aktas {\it et al.}  [H1 Collaboration],
  Eur.\ Phys.\ J.\ C {\bf 36}, 413 (2004)
  [hep-ex/0403056].

\bibitem{Abazov:2007ht} 
  V.~M.~Abazov {\it et al.}  [D0 Collaboration],
  Phys.\ Rev.\ Lett.\  {\bf 99}, 131801 (2007)
  [arXiv:0705.0306 [hep-ex]].

\bibitem{Aaltonen:2009kea} 
  T.~Aaltonen {\it et al.}  [CDF Collaboration],
  Phys.\ Rev.\ Lett.\  {\bf 103}, 021802 (2009)
  [arXiv:0902.1266 [hep-ex]].
\bibitem{Abazov:2011pf} 
  V.~M.~Abazov {\it et al.}  [D0 Collaboration],
  Phys.\ Rev.\ Lett.\  {\bf 108}, 121802 (2012)
  [arXiv:1110.3302 [hep-ex]].

\bibitem{Aad:2012pra} 
  G.~Aad {\it et al.}  [ATLAS Collaboration],
  Phys.\ Lett.\ B {\bf 720}, 277 (2013)
  [arXiv:1211.1597 [hep-ex]].


\bibitem{Aad:2013gva} 
  G.~Aad {\it et al.}  [ATLAS Collaboration],
  Phys.\ Rev.\ D {\bf 88}, no. 11, 112003 (2013)
  [arXiv:1310.6584 [hep-ex]].

\bibitem{ATLAS:2014fka} 
  G.~Aad {\it et al.}  [ATLAS Collaboration],
  arXiv:1411.6795 [hep-ex].
\bibitem{Chatrchyan:2013oca} 
  S.~Chatrchyan {\it et al.}  [CMS Collaboration],
  JHEP {\bf 1307}, 122 (2013)
  [arXiv:1305.0491 [hep-ex]].
\bibitem{cern_note1}  
{\bf CMS} Collaboration, ``CMS at the High-Energy Frontier Contribution to the Update of the European
Strategy for Particle Physics.''
[CMS-NOTE-2012-006]


\end{thebibliography}
\end{document}